\documentclass[sigconf]{acmart}

\AtBeginDocument{%
  \providecommand\BibTeX{{%
    \normalfont B\kern-0.5em{\scshape i\kern-0.25em b}\kern-0.8em\TeX}}}

\usepackage{soul}
\renewcommand\st[1]{}
\usepackage{framed}
\colorlet{shadecolor}{purple!3}
\colorlet{framecolor}{gray!50}

\renewenvironment{shaded*}{%
 \MakeFramed {\advance\hsize-\width \FrameRestore}}%
 {\endMakeFramed}

\copyrightyear{2024}
\acmYear{2024}
\setcopyright{rightsretained}
\acmConference[CHI '24]{Proceedings of the CHI Conference on Human Factors in Computing Systems}{May 11--16, 2024}{Honolulu, HI, USA}
\acmBooktitle{Proceedings of the CHI Conference on Human Factors in Computing Systems (CHI '24), May 11--16, 2024, Honolulu, HI, USA}
\acmDOI{10.1145/3613904.3642849}
\acmISBN{979-8-4007-0330-0/24/05}

\extrafloats{1000}

\begin{document}

\title[The \textit{Situate AI} Guidebook]{The \textit{Situate AI} Guidebook: \\Co-Designing a Toolkit to Support Multi-Stakeholder Early-stage Deliberations Around Public Sector AI Proposals}


\author{Anna Kawakami}
\affiliation{%
  \institution{Carnegie Mellon University}
  \city{Pittsburgh}
  \state{PA}
  \country{USA}
}
\author{Amanda Coston}
\affiliation{%
  \institution{Microsoft Research}
  \city{Cambridge}
  \state{MA}
  \country{USA}
}

\author{Haiyi Zhu}
\authornote{Co-senior authors contributed equally to this research.}
\affiliation{%
  \institution{Carnegie Mellon University}
  \city{Pittsburgh}
  \state{PA}
  \country{USA}
}

\author{Hoda Heidari}
\authornotemark[1]
\affiliation{%
  \institution{Carnegie Mellon University}
  \city{Pittsburgh}
  \state{PA}
  \country{USA}
}

\author{Kenneth Holstein}
\authornotemark[1]
\affiliation{%
  \institution{Carnegie Mellon University}
  \city{Pittsburgh}
  \state{PA}
  \country{USA}
}

\renewcommand{\shortauthors}{Anna Kawakami et. al.}

\begin{abstract}
Public sector agencies are rapidly deploying AI systems to augment or automate critical decisions in real-world contexts like child welfare, criminal justice, and public health. 
A growing body of work \textcolor{black}{documents}\st{has documented} how these AI systems \st{have }\textcolor{black}{often fail}\st{failed} to improve services in practice. These \textcolor{black}{failures}\st{challenges observed downstream} can often be traced to decisions made during \textcolor{black}{the} early \st{problem formulation }stages of AI \textcolor{black}{ideation and }design\textcolor{black}{, such as problem formulation}. However, today, we lack systematic processes to support effective, early-stage decision-making about \textit{whether} and \textit{under what conditions} to move forward with a proposed AI \textcolor{black}{project}\st{innovation}. To understand how to scaffold such processes in real-world settings, we worked with public sector agency leaders, AI developers, frontline workers, and community advocates across four public sector agencies and three community advocacy groups \textcolor{black}{in}\st{situated }\st{across} the United States. Through \textcolor{black}{an} iterative co-design \textcolor{black}{process}\st{sessions}, we created the \emph{Situate AI} Guidebook: a structured \textcolor{black}{process centered around a} set of deliberation questions \textcolor{black}{to scaffold}\st{that scaffolds} conversations \textcolor{black}{around}\st{on} (1) \textcolor{black}{\textit{goals and intended use} \textcolor{black}{for a proposed AI system}, (2) \textit{societal and legal considerations}, (3) \textit{data and modeling constraints}, and (4) \textit{organizational governance factors}.} \textcolor{black}{We}\st{We further} discuss \textcolor{black}{how the guidebook's design is}\st{the process design and success criteria of the guidebook, including how they are} informed by participants’ challenges, needs, and desires for improved deliberation processes. \textcolor{black}{We further elaborate on implications for designing responsible AI toolkits in collaboration with public sector agency stakeholders and opportunities for future work to expand upon the guidebook.} \textcolor{black}{\st{This design methodology}\textcolor{black}{This design approach} can be more broadly adopted to support the \st{utilized in the design}\textcolor{black}{co-creation} of responsible AI toolkits that\st{ and processes aiming to} scaffold key decision-making processes surrounding the use \textcolor{black}{of AI} in the public sector and beyond.}


\end{abstract}





\begin{teaserfigure}
    \includegraphics[scale=0.60]{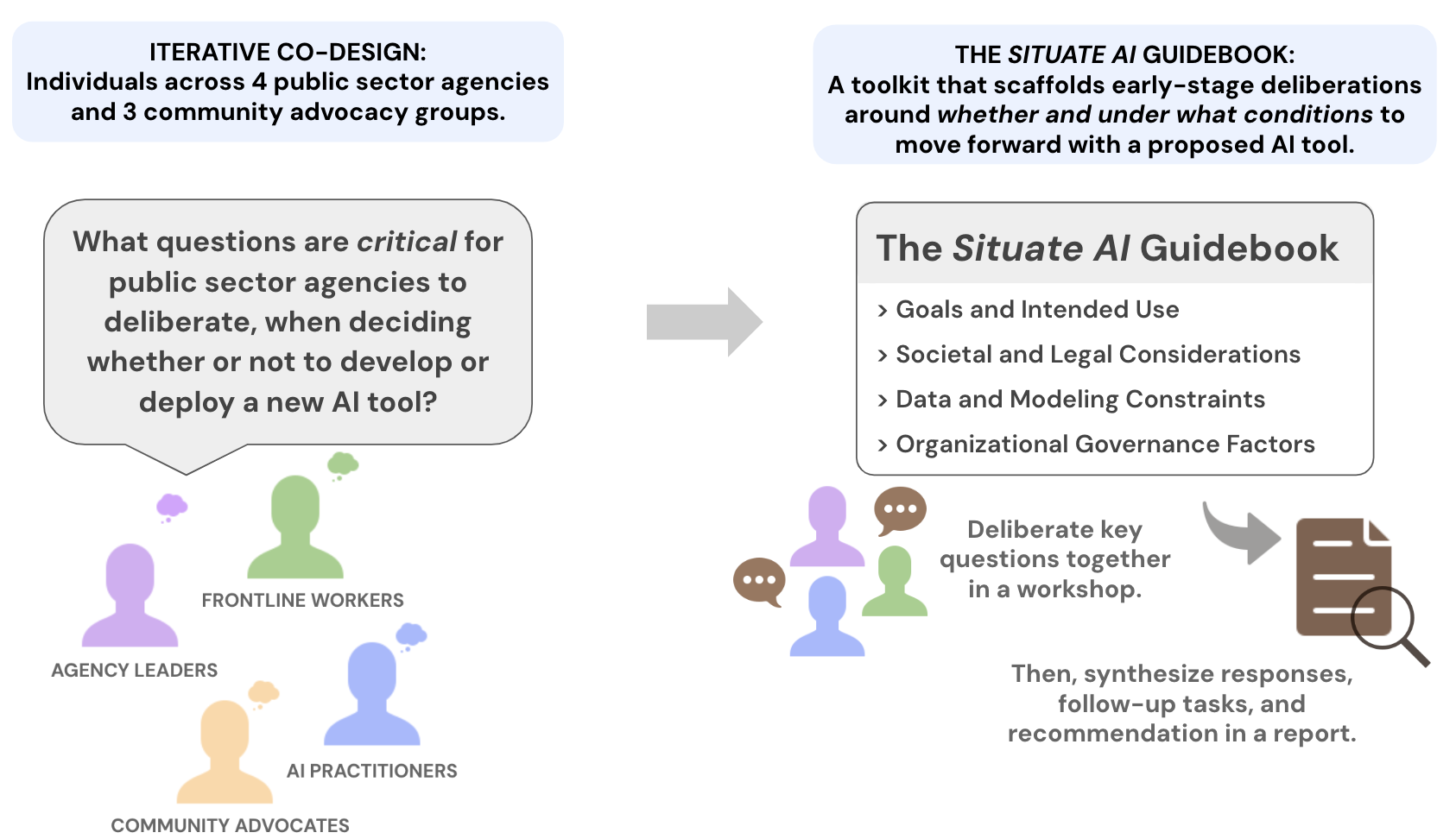}
    \caption{We conducted co-design activities and semi-structured interviews with public sector agency workers (agency leaders, AI practitioners, frontline workers) and community advocates to understand the questions they believed were \textit{critical to discuss} yet currently overlooked before deciding to move forward with a public sector AI proposal. The \textit{Situate AI} Guidebook synthesizes these key considerations into a toolkit to scaffold early-stage deliberations around \textit{whether} and \textit{under what conditions} to move forward with developing or deploying a proposed public sector AI tool.}
     \label{fig:design_process}
  \label{fig:teaser}
\end{teaserfigure}

\maketitle

\section{Introduction}
Public sector agencies \textcolor{black}{in the United States} are rapidly adopting AI systems to assist or automate services in settings such as child welfare, credit lending, housing allocation, and public health. 
\st{\textcolor{black}{In the United States, public sector agencies are government-owned or -affiliated organizations occupying the federal, state, county, or city government, responsible for making decisions around the allocation of educational, welfare, health, and other services to the community~\cite{murray2021public}.}}
\textcolor{black}{These tools have been introduced}\st{Research and development efforts around new AI tools have aimed} to help overcome resource constraints and limitations in human decision-making~\cite{Chouldechova2018, Levy2021}. However, as a growing body of \textcolor{black}{work}\st{research and media \textcolor{black}{scrutiny}} has documented, public sector AI tools\st{deployments} have often failed to produce value in practice, \textcolor{black}{instead exacerbating existing problems or introducing new ones}\st{ introducing additional or amplifying existing problems instead}\st{,\textcolor{black}{, such as exacerbating disparities against marginalized groups.}}~\cite{kawakami2022improving,cheng2022child,wang2022against,narayanan2019recognize}. For example, the Michigan Unemployment Insurance Agency developed an AI-based fraud detection system (MiDAS); the agency stopped using the tool after realizing it falsely flagged over 90\% of its cases---a discovery that was made \textcolor{black}{only} \textit{after} the tool had been in use for over two years, impacting hundreds of thousands of people along the way~\cite{bao2022artificial}. \textcolor{black}{Similarly, following the deployment of an AI-based tool for child maltreatment screening,} Allegheny County's Department of Human Services \textcolor{black}{faced significant criticism after the tool was found to} 
exacerbate biases against Black and disabled communities ~\cite{eubanks2018automating, gerchick2023devil, ho2022algorithm, ho2023algorithm}. \textcolor{black}{Research indicates that these problems in deployment were a consequence of fundamental conflicts between the tool's design on the one hand, and data limitations and worker needs on the other~\cite{eubanks2018automating,gerchick2023devil, kawakami2022improving, kawakami2022care}.} Many other public sector agencies \textcolor{black}{have dropped deployed AI tools for similar reasons, }\st{that have deployed similar AI-based decision support tools have since dropped these tools, }even after investing significant resources in\textcolor{black}{to} their development (e.g.,~\cite{june2022oregon,samant2021family}). 

Many failures in public sector AI projects can be traced back to decisions made during the earliest problem formulation and ideation stages of AI design~\cite{passi2019problem,yildirim2023creating,wang2022against,coston2022validity}. AI design concepts that make it to production may be ``doomed to fail'' from the very beginning, for a variety of reasons. 
For example, AI design concepts have often been conceived in isolation from workers' actual decision-making tasks and challenges, leading to AI deployments that \textcolor{black}{are not actually viable in practice}~\cite{yang2019unremarkable,yildirim2023creating,holstein2017intelligent,kawakami2022care,utterberg2021intelligent}.
\textcolor{black}{Similarly, teams often propose design concepts for new tools that cannot possibly be implemented in an effective, safe, or valid way given technical constraints, such as}
the availability \textcolor{black}{and quality} of data\st{(e.g., unobservable factors are not captured) or quality of data (e.g., data reflects underlying systemic biases)}~\cite{coston2022validity,raji2022fallacy,wang2022against,yildirim2023creating}. \textcolor{black}{However, discussion of  such constraints is commonly left to later \st{project} stages \textcolor{black}{of the AI lifecycle}, by which point teams have invested in an idea and may be more reluctant to explore alternative ideas~\cite{kawakami2022improving,yildirim2023creating}.}
\st{As a result, the problem formulation of these tools make it impossible for them to be implemented in a way that does not conflict with community needs and values around transparent and fair decision-making~\cite{stapleton2022imagining,gerchick2023devil,kawakami2022improving}.}\st{For instance, algorithmic fairness mitigation or ``debiasing'' techniques typically rely upon existing datasets both in order to correct for unfairness and bias, and to assess whether such corrections have been 'successful'. However, when the underlying datasets are severely impacted by harmful biases--as is the case in public sector service contexts--the use of algorithmic debiasing methods can inadvertently amplify the very biases they were intended to correct.}
While \st{public sector} agencies \textcolor{black}{utilizing AI} may be motivated to \textcolor{black}{try to mitigate issues at later project stages, }\st{mitigate demographic disparities or increase performance accuracy in their AI models towards more ``objective'' and ethical decision-making~\cite{dare2017ethical}, }such attempts \textcolor{black}{are unlikely to yield}\st{will not lead to} meaningful improvements if fundamental issues around the problem formulation and solution design are \textcolor{black}{left unaddressed}\st{not first addressed}~\cite{kawakami2022care,yildirim2023creating,stapleton2022imagining,wang2022against,gerchick2023devil}.

In this paper, we ask: \textbf{How can we support public sector agencies in deciding whether or not a proposed AI tool should be \textcolor{black}{developed}\st{designed} and deployed in the first place?} Today, we lack systematic processes to \textcolor{black}{help agencies make informed choices about which \textcolor{black}{AI} project ideas to pursue, and which\st{ideas} are best avoided.}
\textcolor{black}{As AI tools proliferate in the public sector, the failures discussed above indicate}\st{Yet, existing deployments attempts demonstrate} that agencies \textcolor{black}{are repeatedly missing the mark with AI innovation.}\st{may be ``missing the point'' with AI innovation.} 
While existing responsible AI toolkits have provided guidance on ways to support\st{responsible} AI development and implementation \textcolor{black}{to ensure compliance with the relevant principles and values} (e.g., \cite{madaio2020co,mitchell2019model,franzke2021data,richards2020methodology}), most existing toolkits are \textcolor{black}{designed}\st{intended} for \textcolor{black}{use in industry} contexts\textcolor{black}{. Furthemore, most toolkits start from the assumption that the decision to develop a particular AI tool has \textit{already} been made. }\st{and/or}\st{ are designed with the assumption that the }\st{\textcolor{black}{system's development is justified}.}\st{decision to develop \textcolor{black}{a particular AI system has \textit{already} been made}\st{the AI system will remain unquestioned}.}
%

\textcolor{black}{To address these gaps, we introduce}\st{In this paper, we present} the \textit{Situate AI} Guidebook: a toolkit to scaffold early-stage deliberations around \textit{whether and under what conditions} to move forward with the \textcolor{black}{development}\st{implementation} or \textcolor{black}{deployment}\st{adoption} of a \textcolor{black}{proposed}\st{new design concept for a} public sector AI innovation. To ensure that our guidebook and process design \textcolor{black}{is}\st{was} informed by existing organizational needs, practices, and constraints in the public sector, 
we partnered with 32 individuals\textcolor{black}{, spanning a \textcolor{black}{wide} range of roles,} \st{affiliated }across four public sector agencies and three community advocacy groups \st{situated }across the United States. Over the course of 8 months, we iteratively designed and validated the guidebook with a range of stakeholders, including (1) public sector agency leadership, (2) AI developers, (3) frontline workers, and (4) community advocates. 
The public sector agencies we partnered with represent different levels of experience and maturity with\st{decision-making around} AI \textcolor{black}{development and deployment}\st{design and development}: \textcolor{black}{At the time of this research, some had just begun ideating ways to integrate AI tools into their agencies' processes\st{,}\textcolor{black}{;} some were already in the process of developing}\st{Some are already creating} new AI tools\st{,}\textcolor{black}{;} \textcolor{black}{and some had}\st{or have} \textcolor{black}{already} experienced failures in AI tool deployment that led to halts in their use\textcolor{black}{.}\st{; and others are currently (at the time this paper was written) ideating how to integrate AI tools in their agency.} The community advocacy groups include organizations that\textcolor{black}{, among other areas of focus,} represent and support community members \textcolor{black}{in navigating challenging interactions with}\st{that were }\st{negatively impacted by their engagements with}\st{\textcolor{black}{navigating or resisting}}
public services \textcolor{black}{(e.g., parents \textcolor{black}{negatively} impacted by the child welfare system)}\st{ (e.g., foster care system)}. 
  
We conducted formative semi-structured interviews and iterative co-design activities that guided the content and process design of the \textit{Situate AI} Guidebook. In particular\textcolor{black}{:}\st{, the guidebook\textcolor{black}{'s} creation \st{process }involved the following\st{ components}: }
\begin{itemize}
    \item Through \st{the }semi-structured interviews, we \textcolor{black}{developed}\st{formed} an understanding of public sector agencies’ current practices and challenges around \textcolor{black}{the design, development, and evaluation of}\st{ideating, developing, and evaluating} new AI tools, in order to identify \textcolor{black}{opportunities for new processes to improve current practice.}\st{where and how a deliberation-based process can improve their current practices.} 
    \item Through\st{the} co-design activities, participants ideated \textcolor{black}{and iterated upon a set of} questions that they \textcolor{black}{believed were}\st{perceived as} critical to consider before deciding to move forward with \textcolor{black}{the development of a proposed AI tool}.
    \textcolor{black}{In addition, they} 
    described how they envisioned \textcolor{black}{a}
    deliberation process \textcolor{black}{could} 
    be \textcolor{black}{effectively structured for adoption at their agencies.}
\end{itemize}

\textcolor{black}{The resulting set of deliberation questions}\st{Overall, we found that the deliberation questions participants found critical} spanned a broad range of topics, from centering community needs to surfacing \textcolor{black}{potential} agency biases\textcolor{black}{, given}\st{from} their positionality---topics which are relatively understated in existing Responsible AI toolkits developed for industry contexts. \textcolor{black}{\textcolor{black}{Notably, p}articipants gravitated toward deliberation questions that promoted reflection on potential differences in perspective among the various stakeholders of public sector agencies (e.g., agency workers, frontline workers, impacted community members), surrounding topics such as the problem to be solved by an AI tool, notions of ``community'', or understandings of what it means for decision-making to be ``fair'' in a given context.} 
\textcolor{black}{This work presents the following contributions:}

\begin{enumerate}
    \item \textbf{The \textit{Situate AI} Guidebook\footnote{\url{https://annakawakami.github.io/situateAI-guidebook/}} ($Ver.1.0$)} -- the first toolkit \textcolor{black}{co-designed}\st{that is \textit{co-designed}} with public sector agencies and community advocacy groups to scaffold \textit{early-stage deliberations }\st{\textcolor{black}{amongst agency workers}} regarding \textit{whether or not} to move forward with the development of \textcolor{black}{a proposed AI tool}\st{an AI design concept}. 
    \item \textbf{A set of 132 co-designed deliberation questions} spanning \st{across }four high level topics, (1) \emph{goals and intended use}, (2) \emph{societal and legal considerations}, (3) \emph{data and modeling constraints}, and (4) \emph{organizational governance factors}. \textcolor{black}{P}articipants indicated \textcolor{black}{these considerations} are \textit{critical to discuss} when deciding to move forward with \textcolor{black}{with the development of a proposed AI tool,}\st{an AI design concept} yet are not proactively or deliberately discussed today.
   \item \textbf{Guidance on the overall decision-making process} that the \textit{Situate AI} Guidebook can be used to support, informed by how participants envisioned they would use the guidebook in their agencies and \textcolor{black}{by} prior literature discussing related challenges \st{with low adoption rates}\textcolor{black}{\textcolor{black}{that threaten}\st{threatening} the \textcolor{black}{practical} utility} of research-created artifacts~\cite{wong2023seeing,madaio2020co}. 
    \item \textbf{Success criteria for using the guidebook} informed by participants’ existing challenges, prior literature, and signals that participants themselves described as valuable in assessments regarding the guidebook's ability to promote meaningful improvements in their agency. 
\end{enumerate}

\textcolor{black}{\textcolor{black}{In the following sections}\st{For the remainder of the paper}, we \textcolor{black}{first}\st{will} overview \textcolor{black}{relevant bodies of prior}\st{existing} literature \textcolor{black}{to}\st{that} help ground and motivate the creation of our toolkit (Section~\ref{background})\st{,}\textcolor{black}{. We} then describe the \textcolor{black}{approach we took to collaboratively develop the \textit{Situate AI} Guidebook}\st{co-design method we used to create it} (Section~\ref{methods})\textcolor{black}{, and}\st{. We will then} describe \textcolor{black}{the guidebook's major components}\st{the components of the \textit{Situate AI} Guidebook}, including its guiding design principles (Section~\ref{design_principles}), deliberation questions (Section~\ref{content_design}), process design (Section~\ref{process_design}), and success criteria (Section~\ref{indicators}).} \textcolor{black}{We conclude with a discussion of anticipated challenges, as well as directions for future research aimed at understanding how to implement such deliberation processes most effectively\st{ (Section~\ref{discussion})}.} 
\st{The participants from community advocacy groups expressed interest in engaging in deliberation sessions with public sector agencies, in future versions of the \textit{Situate AI} Guidebook that is designed to be used by both agency-internal and agency-external stakeholders.}
We \textcolor{black}{also}\st{additionally} discuss implications for \textcolor{black}{future co-design of}\st{co-designing} responsible AI toolkits intended to promote meaningful change in public sector contexts (Section~\ref{discussion}). The public sector agencies we partnered with in this study \textcolor{black}{plan to}\st{expressed interest in participating in follow-up studies with our team to} explore the use of the guidebook \textcolor{black}{through pilots,}\st{ and} \textcolor{black}{to} identify further avenues for improvement.  

\section{Background}\label{background}
\subsection{Public Sector AI and Overcoming AI Failures}
\textcolor{black}{In the United States, public sector agencies are government-owned or affiliated organizations occupying the federal, state, county, or city government, responsible for making decisions around the allocation of educational, welfare, health, and other services to the community~\cite{murray2021public}.} Public sector agencies across the United States are exploring how to reap the benefits of AI innovations for their own workplaces. AI tools promise new opportunities to improve the efficiency of public sector services, for example, by increasing decision quality and reducing agency costs~\cite{vaithianathan2013children,chouldechova2018case,panattoni2011predictive,billings2012development}. In 2018, 83\% of agency leaders indicated they were willing or able to adopt new AI tools into their agency~\cite{accenture}. In the public sector, there is also a recognition that developing AI tools in-house can help ensure that they are better tailored to meet agency-specific needs, ensure they are trained on representative datasets, and account for local compliance requirements~\cite{engstrom2020government}. However, achieving responsible AI design in the public sector has proven to be an immense challenge~\cite{aclu2021family,veale2018fairness,green2022flaws,green2020false}. The domains \textcolor{black}{where agencies are attempting to apply AI}\st{that agencies are attempting to design or adopt AI tools for} are often highly socially complex and high-stakes–including tasks like screening child maltreatment reports~\cite{saxena2020human}, allocating housing to unhoused people~\cite{kuo2023understanding}, predicting criminal activity~\cite{kleinberg2018human}, or prioritizing medical care for patients~\cite{obermeyer2016predicting}. In these domains, where some public sector agencies have a fraught history of interactions with marginalized communities~\cite{abdurahman2021calculating,roberts2022torn}, it has proven to be particularly challenging to design AI systems that avoid further perpetuating social biases~\cite{cheng2022child}, obfuscating how decisions are made~\cite{kawakami2022care}, or relying on inappropriate quantitative notions of what it means to make accurate decisions~\cite{coston2022validity}. Public sector agencies are increasingly under fire for implementing AI tools that fail to bring value to the communities they serve, \st{and }contributing to a common trend: AI tools are implemented then discarded after failing in practice~\cite{aclu2021family,gravel2006barriers,yang2019unremarkable,yildirim2023creating}. 

Research communities across disciplines (e.g., \textcolor{black}{HCI, machine learning,}\st{theoretical machine learning, HCI,} social sciences, STS) are beginning to converge \textcolor{black}{toward}\st{on} the same conclusion: \textbf{Challenges observed downstream can be traced back to decisions made during early problem formulation stages of AI design}\textcolor{black}{.}\st{, and today} \textcolor{black}{Today}, we lack concrete guidance to support these early stages of AI design~\cite{passi2019problem,wang2022against,coston2022validity,yildirim2023creating}. 
For example, after observing decades of failures to develop clinical decision support tools that bring value to clinicians, researchers have found that AI developers may \textbf{lack an adequate understanding of which tasks clinicians desire support for}, leading to the creation of tools that target problems that clinicians do not actually have~\cite{elwyn2013many,gravel2006barriers,yang2019unremarkable}. These trends are beginning to surface across other domains that have more recently begun to explore the use of new AI tools. 
In social work, researchers have found that technical design decisions in decision support tools reflect \textbf{misunderstandings around the type of work that social workers actually do}, leading to deployments where\textcolor{black}{, for instance,} the underlying logic of the model \st{(e.g., predicting longer-term outcomes) }conflicts with how workers are trained \textcolor{black}{and required} to make decisions~\cite{kawakami2022improving,kawakami2022care}\st{or model outputs (e.g., numerical risk scores) cause confusion for workers who are asked to use them~\cite{kawakami2022care}}. Others have surfaced how seemingly technical design decisions made during early stages of model design \textcolor{black}{actually} embed \textbf{policy decisions that conflict with community values and needs}~\cite{stapleton2022imagining,gerchick2023devil,saxena2023rethinking}. 

In addition to concerns regarding how well the problem formulation and design \textcolor{black}{of}\st{space for} a given AI tool reflects worker practices and community values, there is a concern that AI tools deployed in complex, real-world domains may be \textbf{conceived without adequate consideration for the actual capabilities of AI}. For example, examining a range of real-world decision support tools (e.g., in criminal justice, child welfare, tax lending), researchers have argued that existing AI deployments lack validity, due to limitations in the types of data that can be feasibly collected to train the desired model~\cite{raji2022fallacy, coston2022validity, narayanan2019recognize, wang2022against}. \textcolor{black}{This highlights the need for developers and organizations to reflect upon technical constraints and limitations at earlier stages of the AI development lifecycle, such as when evaluating whether or not to pursue a proposed AI project in the first place.}

While public sector agencies have emphasized the potential to improve decision accuracy and reduce bias as a key motivation to use new AI tools (e.g.,~\cite{dare2017ethical}), these challenges around the problem formulation and design of AI systems implicate the veracity of these claims~\cite{stapleton2022imagining,kawakami2022care,yildirim2023creating,raji2022fallacy,wang2022against,coston2022validity}. 
We identify a significant opportunity to better support public sector agencies in making systematic, deliberate decisions regarding whether or not to implement a given AI \textcolor{black}{tool proposal}\st{design concept}. Given \textcolor{black}{the vast potential for harm, and the similarly vast potential for AI systems to}\st{\textcolor{black}{a}\st{the} vast\textcolor{black}{,}\st{yet} under-explored design space for AI systems \textcolor{black}{to}\st{that could}} meaningfully support workers and improve services in the public sector, it is critical to support agencies through concrete guidance and processes in making more informed decisions around which AI \textcolor{black}{tool proposals}\st{design concepts} to pursue\textcolor{black}{, and which to avoid}. 

\subsection{Toolkits for Responsible AI Governance} 
In an effort to support responsible design and development of AI systems in practice, the HCI, ML and FAccT research communities have contributed a range of responsible AI toolkits. These toolkits are intended to \textcolor{black}{support and} document assessments of AI systems, including their (potential) impacts (e.g., ~\cite{reisman2018algorithmic}), intended use cases (e.g.,~\cite{mitchell2019model}), capabilities and limitations (e.g.,~\cite{richards2020methodology}), dataset quality (e.g.,~\cite{gebru2021datasheets,pushkarna2022data}), and performance measures (e.g., ~\cite{mitchell2019model}). Many of these toolkits are intended to be used as communication tools. For example, Model Cards provide\st{s} a structure for communicating information regarding the intended uses, potential pitfalls, and evaluation measures of a given ML model, to support assessments \textcolor{black}{of suitability}\st{regarding the suitability of an ML model} for a given \textcolor{black}{application and} context of use~\cite{mitchell2019model}. Recent research surveying these toolkits have found that the majority of existing toolkits frame the work of AI ethics as ``technical work for individual technical practitioners\st{,’’ even when beginning with an acknowledgement that responsible AI in itself is an inherently socio-technical concept}~\cite{wong2023seeing}.\st{ In particular, the toolkits target stages within the ML development pipeline, during the AI development, evaluation, deployment, or maintenance stages.} 
For the majority of existing responsible AI toolkits, the primary users are ML practitioners, limiting the forms of knowledge and perspectives that inform the work of ``AI ethics’’~\cite{wong2023seeing}. A smaller number of toolkits have been designed for use by organization-external stakeholders, to support impacted end-users in interrogating and analyzing deployed automated decision systems (\textcolor{black}{e.g.,} the Algorithmic Equity Toolkit~\cite{krafft2021action}); provide impacted stakeholders with an opportunity to share feedback on an AI system's use cases and product design (\textcolor{black}{e.g.,} Community Jury~\cite{azure_2022}); or support philanthropic organizations in vetting public sector AI technology proposals~\cite{vetting_framework}). In all of these examples, the toolkit supports examinations of AI systems that have already been developed and sometimes even deployed. 

\textbf{Most existing toolkits assume that the decision to develop a particular AI system has already been made\st{and/or that the AI system's development is justified}.} 
Therefore, even when they are intended to support reflection and improvement of the AI system, the types of improvements that could stem from using the toolkit \textcolor{black}{tend to be}\st{are} limited to those that would not require \textcolor{black}{fundamental}\st{major} changes to the underlying technology. \textcolor{black}{Meanwhile, while some}\st{Moreover,} existing responsible AI toolkits \textcolor{black}{target}\st{targeting} earlier stages of AI development \textcolor{black}{(e.g.,~\cite{yildirim2023investigating}), these have primarily been designed for}\st{ are primarily intended for } \textit{private sector} contexts. \textcolor{black}{Yet there is good reason to expect that public sector agencies would benefit from tailored responsible AI tools.} \textcolor{black}{For instance, compared}\st{Compared} with the private sector, there is a greater expectation that public sector agencies exist to serve people and are expected to make decisions that center communities' needs. When making decisions as critical as what new AI tools to deploy, agencies are expected to adhere strongly to values such as deliberative decision-making, public accountability, and transparency. \textcolor{black}{To date}\st{Yet}, there exists minimal concrete and actionable guidance on how to support \textit{public sector agencies} in scaffolding \textit{early-stage} deliberation and decision-making. 

A \textcolor{black}{related existing}\st{potentially comparable} artifact is the AI Impact Assessment, described as a ``process for simultaneously documenting an [AI] undertaking, evaluating the impacts it might cause, and assigning responsibility for those impacts''~\cite{moss2021assembling}. 
AI Impact Assessments have been proposed for both public and private sector contexts, and are intended to be completed either at an early stage of AI design (e.g.,~\cite{canada_impact}), or after an AI system is developed or deployed (e.g.,~\cite{Mbuy_Ortolani_2022}). 
\textcolor{black}{Another \textcolor{black}{related}\st{potentially comparable} artifact is the Data Ethics Decision Aid (DEDA)~\cite{franzke2021data}, a framework to scaffold ethical considerations around data projects proposed in the Dutch Government.}
However, \textcolor{black}{neither of these examples are designed as \textit{deliberation toolkits}, to promote collaborative reflection and discussion around the}\st{both of these examples are not designed to prompt \textit{deliberation} and \textit{reflection} on the} underlying problem formulation or solution design of an AI tool. 
\textcolor{black}{AI Impact Assessments and ethical decision aids have}\st{\st{Existing examples online have}\textcolor{black}{They have}} also not \textcolor{black}{typically} been designed in collaboration with the stakeholders they intend to serve. With recent research suggesting \textcolor{black}{low adoption}\st{challenges with low adoption rates} of responsible AI toolkits \textcolor{black}{in real-world organizational contexts}\st{across contexts}~\cite{wong2023seeing,rakova2021responsible}, \textcolor{black}{a co-design approach with organizational stakeholders has the potential to generate responsible AI tools that work in practice.}\st{there remains a critical opportunity to co-design early-stage deliberation toolkits with stakeholders within U.S. public sector agencies--a direction that no research has pursued thus far. }


\section{Methods: Co-Design and Validation of the \textit{Situate AI} Guidebook}\label{methods}
To iteratively co-design and validate the \textit{Situate AI }Guidebook, we conducted semi-structured interviews and co-design activities with a range of stakeholders \textcolor{black}{both}\st{situated} within and outside of public sector agencies. In this section, we describe \st{the }participants’ background\textcolor{black}{s}, the approach and resources \st{we} used in \textcolor{black}{our}\st{the} iterative co-design process, and \textcolor{black}{our data analysis approach.}\st{the analysis method we used.} 

\subsection{Participants and Recruitment}
We co-designed \textcolor{black}{the} \textit{Situate AI} Guidebook with individuals from \textbf{four public sector agencies} across the United States. \textcolor{black}{Collectively}\st{Together}, this set of public sector agencies has experienced a range of decision-making \textcolor{black}{scenarios}\st{situations} around the creation or use of AI-based decision tools. All four agencies are currently ideating new forms of AI-based \st{decision }tools, three have already implemented \textcolor{black}{AI} \st{AI-based decision }tools, and at least one had previously deployed an \textcolor{black}{AI}\st{AI-based decision } tool and subsequently decided to \textcolor{black}{abandon}\st{revoke} it. From these agencies, we wanted to include stakeholders \textcolor{black}{at different levels of}\st{situated across} the organizational hierarchy including those \textcolor{black}{with}\st{that have} experience making relevant decisions and those who are involved in the development or consumption of AI tools but \textcolor{black}{who} are \textcolor{black}{not typically involved in decisions around development and deployment}\st{often left out of decisions}. We therefore included participants from three core stakeholder groups:  1) \textbf{Agency leaders} (\textit{L)} who are in director or managerial roles, typically involved in agency- or department-level decisions including whether to design and deploy a particular AI tool, 2) \textbf{AI developers, analysts, and researchers }(\textit{A}) who are in development, analysis, or research teams internal to a given public sector agency and typically build and evaluate AI tools, and 3) \textbf{Frontline decision-makers} (\textit{F}) whose occupations bring them in direct contact with the community their agency serves and whom an AI tool may be intended to assist. \textcolor{black}{Because we wanted} to learn from additional frontline decision-makers \textcolor{black}{but \textcolor{black}{had access only to a limited number}\st{did not have access to many who were employed} at the public sector agencies we connected with}, we recruited \textcolor{black}{additional}\st{more} participants \textcolor{black}{beyond these agencies, with relevant professional backgrounds.}\st{\textcolor{black}{with relevant background} from academic institutions.} These \textcolor{black}{included}\st{individuals were} social work graduate students with prior field experience making frontline decisions in \textcolor{black}{public sector}\st{social work} agencies. 

\textcolor{black}{In addition, we}\st{We additionally} co-designed the guidebook with individuals from \textbf{three community advocacy groups }across the United States\textcolor{black}{, including}\st{. The three community advocacy organizations are} family representation and child welfare advocacy groups. Individuals from these organizations created the fourth stakeholder group: 4) \textbf{Community advocates} (\textit{C}) who represent and meet community members’ needs around public services. While the Situate AI Guidebook is intended to be used by workers within a public sector agency, we included community advocates because we wanted the guidebook to represent their \textcolor{black}{perspectives regarding}\st{ideas around what} the most critical considerations for moving forward with an AI tool design\st{ includes}. \textcolor{black}{As discussed in Section~\ref{process_comm}, we also worked with community advocates to begin envisioning what a future version of the toolkit, aimed at engaging community members in the deliberation process, might look like.}\st{We additionally solicited guidance on considerations for future versions of the guidebook that may be intentionally (re)designed to also involve community members in the deliberation process. (We discuss these ideas further in Section~\ref{process_comm}).} 

In total, 7 agency leaders; 7 developers, analysts, and researchers; 7 frontline decision-makers; and 11 community advocates participated in the co-design process. \st{Table X includes an overview of participants’ background.}To recruit public sector agencies, we contacted 19 \textcolor{black}{U.S. public sector} agencies \textcolor{black}{at the state, city, or county level with human service\textcolor{black}{s} departments\st{ (e.g., child welfare, predictive policing)}}. We received responses from five agencies. Following a series of informal conversations to share our research goals and study plans, four of the agencies decided to participate in the study. To recruit individuals from community advocacy organizations, we contacted community leaders and advocates across 8 organizations. \textcolor{black}{While we requested individual study participation, some participants preferred to participate in the research study in small groups. By participating in groups, they believed they could provide a more extensive set of insights together\st{, or provide the other participant(s) an opportunity to experience an externally conducted research study}. 21 out of 25 sessions were conducted individually, and the remaining four were group interviews. For ease of communication, we will use the singular noun ``participant'' throughout the remainder of the paper.}

\subsection{Iterative Co-Design and Validation}
The \textit{Situate AI} Guidebook integrates findings \textcolor{black}{across semi-structured interviews and co-design activities, which were conducted}\st{from the semi-structured interview portion of the study, and from the results of the co-design activities. We conducted the semi-structured interviews and co-design activities} over the course of eight months between November 2022 and June 2023. The study sessions were \textasciitilde90 minutes long for public sector workers, who were involved in both the interviews and co-design activities; the study sessions were \textasciitilde60 minutes for community advocates, who were only involved in the co-design activities. 

\subsubsection{Formative Semi-structured Interviews}
To ensure that the \textit{Situate AI} Guidebook is designed to address real-world needs and goals, we conducted semi-structured interviews with public sector agency workers to understand (1) their existing challenges and barriers to making decisions around AI systems and (2) desires for improving their current decision processes. Specifically, \textcolor{black}{to understand existing decision-making processes, we asked \textcolor{black}{each participant} to recall a specific prior experience in which they or their agency needed to decide} 
\st{ we asked \textcolor{black}{each participant} to recall a specific prior experience, to understand how they previously and currently make decisions around}
whether to move forward with the development or use of \textcolor{black}{a new}\st{a given} AI-based \st{decision-making }tool. \textcolor{black}{As participants shared their stories, we asked follow-up questions to}\st{We also asked follow-up questions throughout the interview, to further} probe on possible \textcolor{black}{causes behind the challenges that they described}\st{sources of challenges}. For example, after describing how they previously made a related decision, we asked ``What’s challenging to do well now, when you’re making those decisions?’’ or ``What would you ideally want to discuss in conversations surrounding those decisions?’’ \textcolor{black}{If a participant shared that they had not personally been involved in decisions around AI design and deployment---as was the case with community advocates and many frontline workers---these questions would be skipped, and more time would be spent on discussing these participants' desires for improved decision processes.}\st{Frontline decision-makers and community advocates were not asked these questions, because they were not involved in decisions around AI design and deployment.} \textcolor{black}{We report findings on how agency decision-makers currently make decisions around AI, including how their decisions are shaped by complex power relations they hold with stakeholders external and internal to their agency (e.g., legal systems, frontline workers),} in\textcolor{black}{~\cite{kawakamistudying}}\st{[citation omitted]}. \textcolor{black}{In this paper, we share \st{distinct, }complementary findings that \st{specifically help }provide design rationale for the \textit{Situate AI} Guidebook's \st{deliberation questions and process }design.} 

\subsubsection{Co-Designing the Deliberation Questions} 
\begin{figure}
    \includegraphics[scale=0.35]{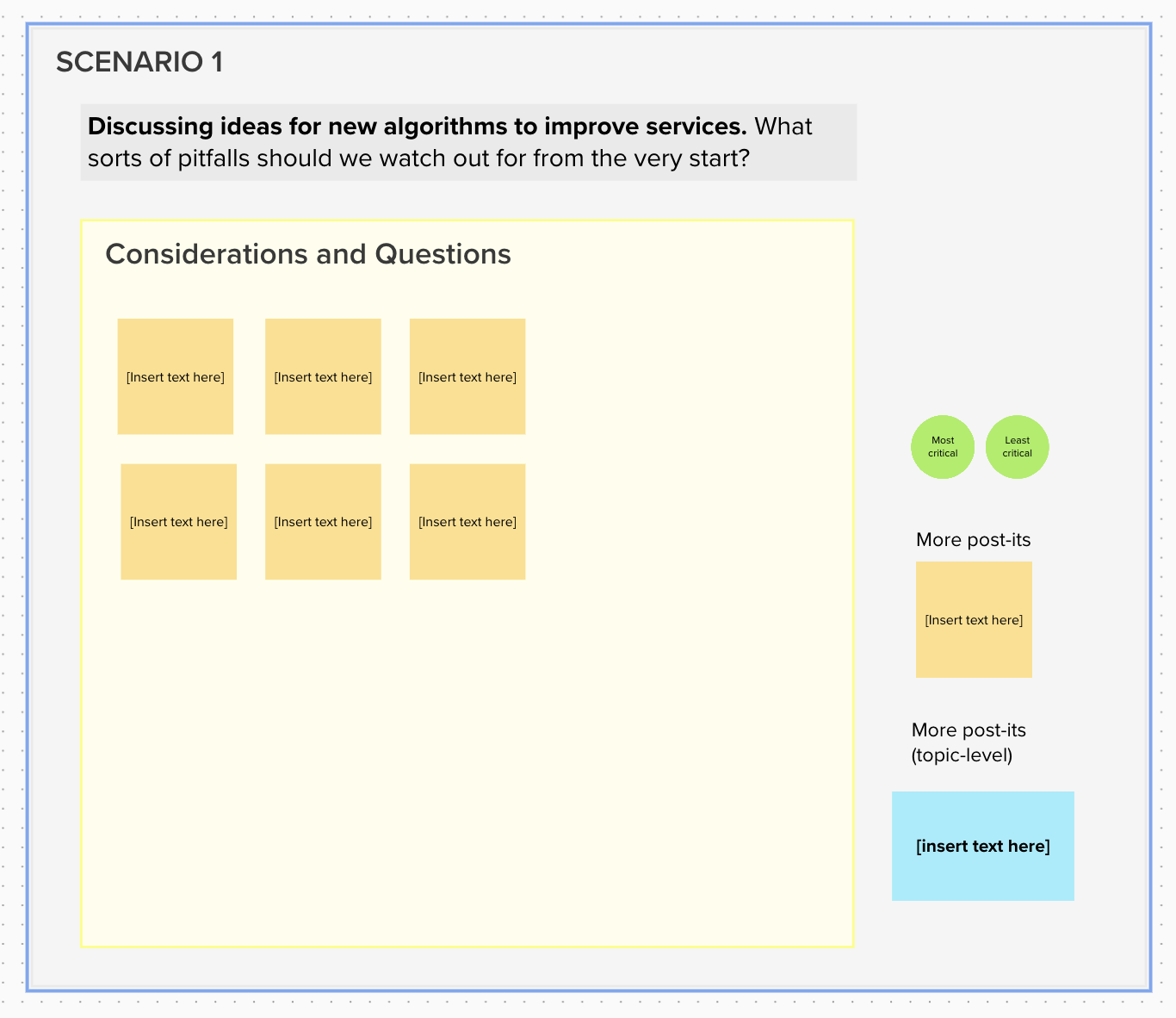}
    \caption{Screenshot of a blank board shown to participants at the start of the co-design activity on Mural. This board presents Scenario 1: Discussing ideas for new algorithms to improve services.}
     \label{fig:figboard}
\end{figure}
In the co-design activity, we first presented \textcolor{black}{each participant} with three potential scenarios: \textcolor{black}{(1)}\st{Scenario 1)} Discussing \textit{ideas} for new algorithms to improve services, \textcolor{black}{(2)}\st{Scenario 2)} Deciding whether to pursue the \textit{development} of a given algorithm design to improve services, and \textcolor{black}{(3)}\st{Scenario 3)} Deciding whether to \textit{adopt} an existing algorithm already implemented by others. We asked \textcolor{black}{the participant} to pick the scenario they had the most experience in or faced the most challenges for. For the scenario they selected, we asked \textcolor{black}{the participant} to think about what critical considerations and questions they believe should be \textcolor{black}{on}\st{at} the table, when deliberating around these scenarios in an ideal future situation. If \textcolor{black}{the participant was} having a challenging time thinking of potential considerations and questions, we provided them with examples \textcolor{black}{that were directly} based on challenges they \textcolor{black}{had} brought up during the semi-structured interview (if applicable). To help document and organize, in real-time, the considerations and questions \textcolor{black}{the participant was} bringing up, we shared our screen and a link to an online board on \textcolor{black}{Mural}\st{Figma}, a collaborative web application where multiple users can \textcolor{black}{generate and arrange sticky notes}\st{engage in interface design}. See Figure~\ref{fig:figboard} for an example of a blank canvas. 

We asked \textcolor{black}{each participant to brainstorm}\st{ to think aloud~\cite{van1994think} the} critical considerations and questions they would want future agencies to discuss. \textcolor{black}{To avoid biasing participants, they were initially asked to openly ideate their own questions without viewing \textcolor{black}{questions}\st{topics} generated by prior participants. Following this, participants were shown existing \textcolor{black}{questions}\st{topics}, providing them an opportunity to comment upon and validate existing questions generated by other participants.} As \textcolor{black}{the participant} openly generated ideas for questions, one of the members of our research team took post-it notes on what they were saying on the \textcolor{black}{Mural}\st{Figma} board. \textcolor{black}{The researcher}\st{We} would frequently check in with the participant, to ensure the post-its accurately represented their ideas. We also welcomed them to edit the post-its or create new ones. As they brainstormed, we asked follow-up questions to better understand how they think a given question could get answered, what makes it challenging to answer the question now, or how they are conceptualizing certain terms. For example, when a participant generated the question ``How well are we involving community members in these decisions?,’’ we asked them to further elaborate on what this might look like in practice. This generated additional post-its, like ``How well do we understand the costs, risks, and effort required of community members, if we invite them to contribute to model design decisions?’’ and ``How are we weighting false positives and false negatives in a given algorithm, based on what type of mistake that is for the impacted community members?’’ 

\textcolor{black}{As mentioned above, after}\st{After} \textcolor{black}{the participant} generated their own questions and considerations on the blank canvas, they were shown a list of topics and example questions for additional consideration. \textcolor{black}{This helped scaffold further ideation on any considerations they may have missed in their initial ideation.} In the first study \textcolor{black}{session}, we provided an initial list of eight broad topics, informed by prior literature: (a) Overall goal for using algorithmic tool, \textcolor{black}{(b) Selection of outcomes that the algorithmic tool should predict, (c) Empirical evaluations of algorithmic tool, (d) Legal and ethical considerations around use of algorithmic tool, (e) Selection of training data for algorithmic tool, (f) Selection of statistical models to fit data, (g) Long-run maintenance of algorithmic tool, and (h) Organizational policies and resources around use of algorithmic tool}. We prompted \textcolor{black}{the participant} to discuss any new ideas the provided topical categories inspired, or any disagreements they had with the categories. Figure~\ref{fig:design_process} shows an example of what this list looked like in \textcolor{black}{later}\st{more mature} stages of the co-design process. 

\textcolor{black}{Between study sessions}\st{After each research session, and before the next one,} one or more researchers in our team iterated on the post-its generated during that study, to reduce redundancies and improve clarity. We then grouped the individual questions and considerations underneath the existing topical categories, \textcolor{black}{while} iteratively \textcolor{black}{refining categories or} creating new categories and subcategories as needed. The next participant was shown this updated version of the aggregated questions and topics at the end of the study. 

\begin{figure*}
    \includegraphics[scale=0.45]{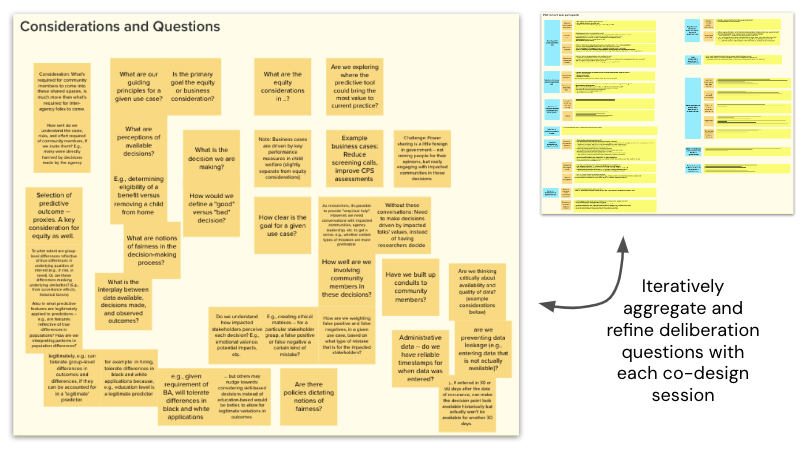}
    \caption{Screenshot of a Mural board populated with post-its after one participant's co-design activity. In our iterative co-design process, these post-its were refined by the research team then added to an aggregated list of questions that were successively grouped into higher level categories. }
     \label{fig:design_process}
\end{figure*}

\subsubsection{Guidebook Reflection and Validation}
Participants that contributed to later stages of the co-design \textcolor{black}{process}\st{sessions} were shown an overview of the aggregated questions, a recommended deliverable for the deliberation guidebook, and a high-level outline of a proposed deliberation process\st{that the generated deliberation questions can be used within}. We first showed \textcolor{black}{participants} the aggregated questions, and asked if there were any questions that they felt were critical to include but missing. We additionally asked if they disagreed with the importance of any \textcolor{black}{of the} question\textcolor{black}{s}, or if the wording of a\textcolor{black}{ny} \st{particular }question was confusing in any way. We then showed the participant an overview of the deliberation \textcolor{black}{process} and asked for their perspectives around what they would like to change in the proposed process, to have it fit better into their existing organizational decision-making process\textcolor{black}{es}. To address challenges in the potential use of the protocol, we also asked \textcolor{black}{participants} (especially frontline workers and community advocates) about challenges they anticipate with participating in the deliberation \textcolor{black}{process. We \textcolor{black}{then} invited participants to discuss potential adjustments to the process}\st{protocol, including interventions} or alternative processes that can help \textcolor{black}{address these challenges and} create a safer environment for them.

\subsection{Qualitative Analysis}
\textcolor{black}{The study recordings from the semi-structured interviews and co-design activities were transcribed and then qualitatively coded \textcolor{black}{by two members of the research team} using \textcolor{black}{a} reflexive thematic analysis \textcolor{black}{approach}~\cite{braun2019reflecting}\st{ by two researchers}. We ensured that all interviews were coded by the first author, who conducted all of the interviews and, whenever applicable, another author who observed the interview. The first author coded one transcript first, then discussed the codes with other \textcolor{black}{coder}s to align on coding granularity. Each coder prioritized coding underlying reasons why participants generated certain questions during the co-design activity, while also remaining open to capturing a broader range of potential findings. We resolved disagreements between coders through discussion.}
%

\section{The \textit{Situate AI} Guidebook}
The Situate AI Guidebook is a process to scaffold early-stage deliberations around \textit{whether and under what conditions} to move forward with the \textcolor{black}{development or deployment of a proposed}\st{implementation or adoption of a new}\st{idea for a public sector} AI innovation. The current version of this toolkit is intended for use within public sector agencies at various stages of maturity in their use of AI tools---from those that are just beginning to consider the use of new AI tools to those that may already have years of experience deploying AI tools. The deliberation questions are designed to be discussed across different stakeholders employed in a public sector agency\textcolor{black}{, such as}\st{:} agency leadership, AI practitioners and analysts, program managers, and frontline workers. 

In this section, we provide an overview of the Situate AI Guidebook as an outcome of our co-design and validation sessions. We describe the Situate AI Guidebook through the following sections: (Section~\ref{design_principles}) Guiding Design Principles, (Section~\ref{content_design}) Content Design, (Section~\ref{process_design}) Process Design, and (Section~\ref{indicators}) Success \textcolor{black}{Criteria} for Use. 

To provide context for key design decisions, throughout each section, we elaborate on participants’ existing practices, challenges, desires, and needs for improving their decision-making process\textcolor{black}{, drawing upon our thematic analysis}. Where appropriate, we describe how the Situate AI Guidebook compares with existing responsible AI toolkits. At times, participants diverged in their desires (e.g., regarding how the decision-making process should be integrated into their agency). In some of these cases, our research team integrated these disagreements into the design of the guidebook (see Design Principle 2); in other cases, we document how these disagreements \textcolor{black}{present challenges for the use of the guidebook, suggesting}\st{may challenge the use of the guidebook and present themselves as} opportunities for future work (Section~\ref{process_design} and Section~\ref{discussion}). 

\subsection{Guiding Design Principles} \label{design_principles}
The goal of the guidebook is to scaffold public sector agency decision-making around the following question: \textit{Should we move forward with \textcolor{black}{developing or deploying a proposed AI tool}\st{implementing the AI tool design concept}? If yes, what are key considerations to plan for?} The guidebook aims to support agencies in answering this question through a deliberation-driven process \st{involving}\textcolor{black}{\textcolor{black}{supported}\st{constructed} by the following materials}: (1) Question prompts to support conversations around the social (organizational, societal, and legal) and technical (data and modeling) considerations that should \textcolor{black}{inform}\st{provide supporting evidence for} their recommendation, (2) Pointers to external resources to help guide their responses, (3) Template for a recommended deliverable to help \textcolor{black}{communicate rationales and}\st{formulate} evidence for \textcolor{black}{the recommendation that results from these deliberations}\st{their recommendation based on the deliberations}, (4) Proposed use cases that illustrate how agencies could adopt the guidebook into their existing work processes, and (5) Success \textcolor{black}{criteria} to signal whether the intended outcomes of the guidebook may be relevant and useful to agencies. 

In co-designing the guidebook towards this goal, we centered two core design principles: 
\begin{itemize}
    \item \textbf{(Design Principle 1) Promoting reflexive deliberation.} The question prompts (Section~\ref{content_design}) should support \textcolor{black}{stakeholders}\st{users} in having \textcolor{black}{reflexive discussions}\st{deeper deliberations}---for example, conversations that surface their own pre-existing assumptions and beliefs about \textcolor{black}{human versus} AI capabilities and limitations \textcolor{black}{with respect to a given task and context}, or \textcolor{black}{that surface}\st{communicating} relevant tacit knowledge that may be helpful to share with others. The question prompts should be designed to avoid prompting simple yes or no responses\textcolor{black}{, to ensure that responses to complex questions are not reduced to a simple compliance activity}. In drawing on prior work that emphasize the role of the toolkit as one that ``prompts discussion and reflection that might not otherwise take place''~\cite{madaio2020co,shilton2013values}, this design principles extends these notions of effective toolkits from prior literature to apply to topics of importance in public sector contexts. \textcolor{black}{Prior research on public sector contexts (e.g.,~\cite{HoltenMoller2020,kawakami2022improving,stapleton2022imagining}) as well as findings from this study suggest that agency stakeholders' differing backgrounds shape their assumptions and concerns around AI tools, motivating the need for a deliberative decision-making process that surface these individual differences.} Throughout Section~\ref{content_design}, we elaborate on participants’ existing challenges and desires to illustrate the importance of Design Principle 1 in their contexts. 
    \item \textbf{(Design Principle 2) Ensuring practicality of the process.} The guidebook should be designed to support a process (Section~\ref{process_design}) that public sector agencies can feasibly understand, adopt, and adapt as needed. If an agency already has an existing decision-making structure, or conversations related to AI design already take place, the agency should find \textcolor{black}{it easy to ``fit'' the guidebook into their existing organizational processes and conversations}\st{the task of ``fitting'' the guidebook into their existing organizational processes and augmenting their existing conversations a practical and feasible one}. This design principle \textcolor{black}{is aimed at}\st{intends to ensure that the guidebook} addresses\textcolor{black}{ing} concerns \textcolor{black}{raised in}\st{from} prior literature \textcolor{black}{that}\st{critiquing how} existing responsible AI toolkits are often designed in isolation from the organizational contexts they intend to augment (e.g.,~\cite{wong2023seeing}). \textcolor{black}{This design principle is also motivated by our observations of the four public sector agencies in our study, which each had their own existing or \textcolor{black}{planned}\st{envisioned} organizational processes for developing AI tools (Section~\ref{process_design_usecase}).} Further, by co-designing the \textit{process} the toolkit should follow (in addition to the guidebook content), we \st{additionally }further an understanding of how organizational, labor, and power dynamics implicate the potential effectiveness of responsible AI toolkits in the public sector \textcolor{black}{(Section~\ref{process_design_power})}. 
\end{itemize}

\subsection{Content Design: Scaffolding Reflexive Deliberation } \label{content_design}
Participants ideated critical questions that spanned four high-level topics, 12 mid-level, and 20 low-level topics.\st{Table X includes a high-level overview of the section and subsection names.} In each of the four subsections below, we briefly describe why participants were interested in the overall category of questions and provide example questions. The full set of deliberation questions for the \textit{Situate AI} Guidebook ($Ver.1$) can be found in Appendix \ref{appendix}.
\subsubsection{\textbf{Goals and Intended Use}} \label{topic_goals}
\textbf{}
\begin{shaded*}
This section is intended to scaffold conversations around the following broad questions: 
\begin{enumerate}
    \item Given our underlying goals and intended use case(s), is our proposed AI tool appropriate? 
    \item What evidence \st{currently exists to }\textcolor{black}{do we have to support our answer to the previous question? What}\st{suggest your response is true, and what} additional tasks may be required in the future to help \textcolor{black}{us}\st{you} gather more evidence \textcolor{black}{and/or better understand the evidence we currently have}\st{/form a better understanding of the question responses}? 
\end{enumerate}
\paragraph{Sample Questions}
\begin{itemize}
    \item Overall goal for using algorithmic tool
    \begin{itemize}
        \item Who is going to be affected by the decision to use this hypothetical AI tool? 
        \item What evidence \textcolor{black}{do we have suggesting that the painpoint this tool aims to solve actually exists?}\st{exists to suggest a pain point this tool aims to solve}? What evidence \textcolor{black}{do we have}\st{exists to} suggest\textcolor{black}{ing that} technology may offer a remedy to this pain point? 
        \item Recall the \textcolor{black}{stakeholders who are the most impacted by this hypothetical AI tool}. How do we \textcolor{black}{bring their voices to the table when determining goals?}\st{get the voices of those most impacted to the table, for example, to ensure we envision all foreseeable impacts?} 
        \item Are there differences in the goals the agency versus community members think the tool should address? If so, what are they? If \textcolor{black}{we}\st{you} are uncertain, what \textcolor{black}{can we do to understand}\st{are your plans for understanding} potential differences? 
        \item What biases (as a \textcolor{black}{public sector} agency) do we bring into this decision-making process? 

    \end{itemize}
    \item Selection of outcomes that the algorithmic tool aims to improve 
    \begin{itemize}
        \item \textcolor{black}{Hypothetically, imagine that our tool does a perfect job of improving the outcome that it targets. What}\st{How might solving the tool's target problem create} additional problems \textcolor{black}{might this create} elsewhere in the system?
    \end{itemize}
    \item Empirical evaluations of algorithmic tool 
    \begin{itemize}
        \item Once the tool is deployed and in use, how can we evaluate how well it is working in the short-term? How can we evaluate how well it is working longer-term? 
        \item \textcolor{black}{How can we effectively evaluate}\st{Are we assessing} the tool from the perspective of impacted community members? \st{What evidence do we have to suggest that we are genuinely understanding their concerns and desires?} 
        \item How might frontline workers respond to the tool? How can we better understand their underlying concerns and desires towards the tool? 

    \end{itemize}
\end{itemize} 
\end{shaded*}

The deliberation questions focus on promoting conversations that bridge reflection and understanding of the goals of the proposed AI tool, as well as how these goals will be operationalized into measurable outcomes. \textcolor{black}{The 52} questions within the \textit{Goals and Intended \textcolor{black}{U}\st{u}se} section are divided into nine subsections: (1) Who the tool impacts and serves, (2) Intended use, (3) How agency-external stakeholders should be involved in determining goals, (4) Differences in goals between the agency and impacted community members, (5) Envisioned harms and benefits, (6) Impacts of outcome choice, (7) Measuring improvement based on outcomes, (8) Centering community needs, and (9) Worker perceptions. 
For the purpose of this paper, we sample one question from each topical subsection.

Several of the questions in this section are designed to help \textbf{surface underlying assumptions regarding who benefits} from the use of the tool, and \textcolor{black}{to support discussion around} \textbf{what evidence suggests that these assumptions are true}. These questions stem from participants’ concerns around whether their AI systems are targeting areas that would bring the most benefits, and to whom these benefits apply. 
For example, one participant noted that their agency had invested a lot of effort into assessing and trying to improve fairness in their algorithms. However, the participant wondered whether they should have been having conversations around larger, ``more challenging'' questions. For instance, they wondered whether ``correcting for bias'' in an algorithm within an inherently biased system is a meaningful or feasible goal. They further elaborated: 
\begin{quote}
    ``I think there my concern often has to do with [the] unexamined belief that an algorithm is always an improvement. [...] I think [questions on broader goals and benefits are] more challenging and that people [who] are running the system may not always see [...] Personally, I think there's a lot of stuff that can be done with machine learning that doesn't have to [\textcolor{black}{target}\st{... program …}] decision-making at the participant level. [...] But those are the kinds of questions the immediate focus [is] on. `Oh, we're going to use this to make decisions at critical points in programs.' Those are things that to me still need to be discussed. And it may be that those conversations are happening at tables that I'm just not at.'' (A02) 
\end{quote}

Other participants expressed concerns for how frontline workers in their agency–the majority of who are currently not involved in early-stage conversations around the goals of the AI tool–may be \textbf{misunderstanding the intended uses and capabilities of their AI tools}. For example, one participant described that frontline workers may be concerned that the AI tools will displace them, even though their agency doesn’t intend to use them to automate workers’ jobs. They described:
\begin{quote}
    “There's almost like a mystique around machine learning algorithms, like there's some amazing thing that is all knowing and all seen, and therefore can predict all these different things. [...] helping people [... understand] what it's able to do and not able to do, I think, is something we've struggled with” (A04). 
\end{quote}
Other questions are intended to help \textbf{forefront considerations around what additional planning and resources may be needed}, in order to adequately complete a related task in the future. For example, workers within agencies often described that involving community members in their AI design and evaluation process can be challenging, given the current lack of infrastructure to support such collaborations. However, community advocates described how involving community members is often an after-thought. One community advocate described the importance of being intentional and proactive in community engagement practices, because 
\begin{quote}
    “it's easy to let that be something that gets back burning, like throughout the process to just have that be something we'll get to, and then we end up in that feedback loop where the feedback is provided but the tool is already created” (C2). 
\end{quote}
Questions in this section help promote earlier reflection and planning on how community members could be involved, so that they could conduct appropriate empirical evaluations regarding their perceptions of the AI tool. 


\subsubsection{\textbf{Societal and Legal Considerations}} \label{topic_social}
\textbf{}
\begin{shaded*}
This section is intended to scaffold conversations around the following broad questions: 
\begin{enumerate}
    \item \textcolor{black}{Given the societal, ethical, and legal considerations and envisioned impacts associated with the use of AI tools for our stated goals,}\st{Given the context-specific societal and legal factors that impact, and are impacted by, the design of this proposed AI tool,} is our AI tool appropriate?
    \item What evidence \st{currently exists to }\textcolor{black}{do we have to support our answer to the previous question? What}\st{suggest your response is true, and what} additional tasks may be required in the future to help \textcolor{black}{us}\st{you} gather more evidence \textcolor{black}{and/or better understand the evidence we currently have}\st{/form a better understanding of the question responses}?
\end{enumerate}

\paragraph{Sample Questions}
\begin{itemize}
    \item Legal considerations around the use of algorithmic tool 
    \begin{itemize}
        \item Do the people impacted by the tool have the power or ability to take legal recourse?
    \end{itemize}
    \item Ethical and fairness considerations around the use of algorithmic tool 
    \begin{itemize}
        \item Are there differences in the goals the agency versus community members think the tool should address? If so, what are they? If you are uncertain, what are your plans for understanding potential differences? 
        \st{\item What are underlying assumptions that tool developers/researchers may have, regarding the soundness of the design decisions made in the tool?}
        \item Can we agree on a definition of fairness and equity \textcolor{black}{in this context}? \textcolor{black}{What}\st{How} would it look like if the desired state is achieved?
        \item Are \textcolor{black}{fairness and equity}\st{bias} definitions and operationalizations adequately context-specific? (For example, in the child welfare domain: children with similar profiles receive similar predictions irrespective of race?)
        \item Do we understand the negative impacts of the decision made across sensitive demographic groups?

    \end{itemize}
    \item Social and historical context surrounding the use of algorithmic tool 
    \begin{itemize}
        \item Have \textcolor{black}{we}\st{you} recognized and tried to adjust for implicit biases and \textcolor{black}{discrimination}\st{racism} inherent in these social systems that might get embedded into the algorithm?
        \item How might \textcolor{black}{we clearly}\st{you adequately} communicate the limitations and historical context of the data to community members?

    \end{itemize}
\end{itemize}
\end{shaded*}

Overall, the goal of this section is to help promote a systematic, deeper conversation on the various dimensions of social and ethical concerns relevant to the design of an AI tool. \textcolor{black}{The 38} questions within the \textit{Societal and Legal Considerations} section are divided into seven subsections: \textcolor{black}{(1)} Legal considerations around the use of the algorithmic tool, \textcolor{black}{(2)} Impacted community member needs, \textcolor{black}{(3)} Involving impacted communities, \textcolor{black}{(4)} Clarity of ethics goals and definitions, \textcolor{black}{(5)} Operationalization of ethics goals, \textcolor{black}{(6)} Envisioning potential negative impacts, and \textcolor{black}{(7)} Social and historical context surrounding the use of the algorithmic tool. Again, for the purpose of the paper, we sample one question per topical subsection. 

Participants \textcolor{black}{shared that} they did not currently have \textbf{structured opportunities to proactively discuss social and ethical considerations surrounding} AI tool design. While participants described that their teams spent a lot of time working on related data- and model-specific fairness tasks (e.g., using bias correction methods to improve the fairness of their AI tool), several participants noted a \textbf{desire to discuss normative concerns regarding the design of an AI tool that could only be addressed in earlier problem formulation stages}. 
Moreover, participants’ past experiences illustrated an opportunity to better support cross-stakeholder communications around the ethical considerations that should aid AI design, by equipping teams with a \textbf{shared knowledge base and vocabulary} for ethical concerns. For instance, one participant described how a leadership team tasked them with creating a predictive algorithm to assist decisions about fraud investigation. The participant’s team tried to ``get them away from this'' because the task was technically infeasible (producing high false positive rates) and ethically risky the cost of errors is high, given that decisions to investigate are highly intrusive to the individual\st{)}. This section's questions intend to support agency stakeholders in forming a more complete understanding of the different ethical factors that could make a proposed AI tool design ``appropriate'' or ``inappropriate.'' 

We note that the guidebook does not exclusively surface societal and ethical considerations in this section; the prevalence of relevant questions included in the other three topical sections (Goals and Intended Use, Data and Modeling Constraints, Organizational Governance) reflect how social and ethical considerations are intertwined with all facets of a proposed AI tool. 

\subsubsection{\textbf{Data and Modeling Constraints}} 
\textbf{}
\begin{shaded*}
This section is intended to scaffold conversations around the following broad questions: 
\begin{enumerate}
    \item Given the availability and condition of existing data sources, and our intended modeling approach, is our proposed AI tool appropriate? 
    \item What evidence do we have to support our answer to the previous question? What additional tasks may be required in the future to help us gather more evidence and/or better understand the evidence we currently have?
\end{enumerate}

\paragraph{Sample Questions}
\begin{itemize}
    \item Understanding data quality
    \begin{itemize}
        \item Has the definition of the data changed over time? (E.g., in child welfare, has reunification always meant to reunify with the parent?)

    \end{itemize}
    \item Process of preparing data
    \begin{itemize}
        \item How are we preprocessing the data?
        \item Who should be involved in making decisions around whether to include or exclude certain data points or features? Do we have plans for involving those people? 

    \end{itemize}
    \item \textcolor{black}{Model selection}\st{Selection of statistical models to fit data} 
    \begin{itemize}
        \item Is our model appropriate given the available data\st{(e.g., tabular data with outliers)}? \textcolor{black}{Why or why not?}
    \end{itemize}
\end{itemize}
\end{shaded*}

\textcolor{black}{This section intends to forefront conversations around data and technical work that may be critical to have earlier on. \textcolor{black}{The 18} questions within the \textit{Data and Modeling Constraints} section are divided into seven subsections: \textcolor{black}{(1)} Understanding data quality, \textcolor{black}{(2)} Process of preparing data, \textcolor{black}{(3)} Selection of statistical models to fit data. For the purpose of the paper, we provide a subsample of questions under each topical subsection.}

Participants who had experience developing AI tools often underscored the importance of ensuring that they had the computing resources and data needed to develop their proposed AI tool. For example, they described the importance of forming a context-specific understanding of the data labels that may be challenging to identify without relevant domain knowledge (e.g., whether certain labels like “reunification” have changed definitions over time). Others described the importance of deliberating who should be involved in data inclusion and exclusion decisions when they are cleaning their data. 

\subsubsection{\textbf{Organizational Governance Factors}} \label{topic_org}
\textbf{}
\begin{shaded*}
This section is intended to scaffold conversations around the following broad questions: 
\begin{enumerate}
    \item Given our plans for ensuring longer-term technical maintenance and policy-oriented governance, do we have adequate post-deployment support for our proposed AI tool?
    \item What evidence \st{currently exists to }\textcolor{black}{do we have to support our answer to the previous question? What}\st{suggest your response is true, and what} additional tasks may be required in the future to help \textcolor{black}{us}\st{you} gather more evidence \textcolor{black}{and/or better understand the evidence we currently have}\st{/form a better understanding of the question responses}?
\end{enumerate}

\paragraph{Sample Questions}
\begin{itemize}
    \item Long-run maintenance of algorithmic tool 
    \begin{itemize}
        \item Do we \textcolor{black}{expect}\st{believe} there will be shifts in performance metrics over time? \textcolor{black}{If so, why?} What are our plans for identifying and mitigating those shifts? 
        \item \textcolor{black}{Do we have the mechanisms to monitor whether the tool is having unintended consequences?}

    \end{itemize}
    \item  Organizational policies and resources around the use of algorithmic tool
    \begin{itemize}
        \item Is there training for frontline workers who will be asked to use the tool? What evidence suggests that this training is adequate? 
        \item \textcolor{black}{Imagine that we could assemble the ``ideal team'' to monitor and govern the tool after it is deployed: What are the characteristics of this ideal team?}\st{What are the characteristics of the most ideal group that could stay on after this tool is deployed, to help monitor and govern the tool? }
        \begin{itemize}
            \item \textcolor{black}{Who is the \textit{actual} team that will monitor and govern the tool after it is deployed?}\st{What are the characteristics of the actual group that will stay on after the tool is deployed? What new risks to post-deployment monitoring and governance may be introduced by deviations from the most ideal case?}
            \item \textcolor{black}{Given the gaps between the “ideal team” and the actual team we expect to have: What risks to post-deployment monitoring and governance can we anticipate? How might we mitigate these risks?}
        \end{itemize}

    \end{itemize}
    \item Internal political considerations around the use of algorithmic tool
    \begin{itemize}
        \item How well do we understand system administrators' and leadership's perspectives around the use of this tool?
        \item How well do staff and leadership understand `why' the tool could bring value?

    \end{itemize}
\end{itemize}
\end{shaded*}
 \textcolor{black}{The 24 }questions within the \textit{Organizational Governance Factors} section are divided into five subsections: \textcolor{black}{(1)} Measuring changes in model performance over time, \textcolor{black}{(2)} Mechanisms to identify long-term changes in model performance, \textcolor{black}{(3)} Policies around worker interactions with the AI tool, \textcolor{black}{(4)} Governance structures around the AI tool, and \textcolor{black}{(5)} Internal political considerations around the use of the AI tool. As with prior sections, we include in this paper a sample of questions across these topical subsections. 

Similar to considerations around the \textit{Social and Legal Considerations} of AI design (Section~\ref{topic_social}), participants often described encountering challenges when attempting to meet organizational governance-related needs of the AI tool, like maintaining their AI tool over time, ensuring workers are adequately trained, or communicating the goals and capabilities of the AI tool to agency leadership. \textcolor{black}{Partipants highlighted that many of these challenges arise because}\st{Oftentimes, participants described challenges from having} \textcolor{black}{such considerations are discussed}\st{these considerations discussed} \textcolor{black}{in an ad-hoc manner,} too late \textcolor{black}{in the AI development process.}\st{into their AI development cycle and in an ad-hoc manner.} Given \textcolor{black}{that} several of these needs may require longer-term planning and preparation (e.g., gathering resources of model maintenance), public sector agencies may be better equipped in meeting these governance needs if they were discussed in early stages of model design (rather than after \textcolor{black}{an}\st{the} AI tool \textcolor{black}{has already been}\st{is already} developed). For example, participants described how they currently lack domain experts that could \textcolor{black}{help} maintain and improve their model post-deployment---a gap in their AI development process that they felt was critically important to address. While agencies \st{may }currently discuss maintenance-related concerns \textcolor{black}{at the deployment stage}\st{during the implementation stages of their AI development cycle}, this may not allow the agencies enough time to deliberate who should be involved in maintenance, or how to allocate additional roles for a maintenance team. 

\subsection{Process Design: Designing for Practicality and Adaptability } \label{process_design}
The overall goal of the Situate AI Guidebook is to help public sector agencies make more informed, deliberative decisions about whether and how to move forward with implementing a proposed AI tool. Prior literature studying existing responsible AI toolkits have started to surface concerns around how such toolkits may be \textcolor{black}{used inappropriately}\st{inappropriately used} or not used at all in practice, due to misalignments with the organizational contexts \textcolor{black}{they are}\st{it is} designed to support~\cite{wong2023seeing,madaio2020co}. In this section, we describe findings related to the broader deliberation \textit{process} that participants envisioned the deliberation questions (Section~\ref{content_design}) could be used to support. 

Below, we first present our proposed use case for the \textit{Situate AI} Guidebook, \textcolor{black}{along with}\st{including} an example instantiation of the use case and \textcolor{black}{an explanation}\st{explanations} of how participants' existing practices informed \textcolor{black}{this}\st{the} use case. We then discuss participants' desires for alternative use cases and processes around deliberation. Participants across agencies and roles expressed interest in \st{additionally }using the questions in a few different ways, based on their concerns around cross-stakeholder power dynamics and desires to enable deliberation practices that align with their organizational values~\cite{wong2023seeing,rakova2021responsible}. \textcolor{black}{Given}\st{Due to} participants' interests in adapting the guidebook to different use cases, a key component of the \textit{Situate AI }Guidebook is that it is designed to allow users to select which topics and questions they would like to focus on: The deliberation questions are categorized and grouped into modular\st{, movable} components; and users have the flexibility to select from a large set of deliberation questions within each component to identify a subset that is most relevant to their use case.

\subsubsection{\textbf{Proposed Use Case: Using the Guidebook to Support Structured and Iterative Deliberations}} \label{process_design_usecase}

\begin{figure*}
    \includegraphics[scale=0.3]{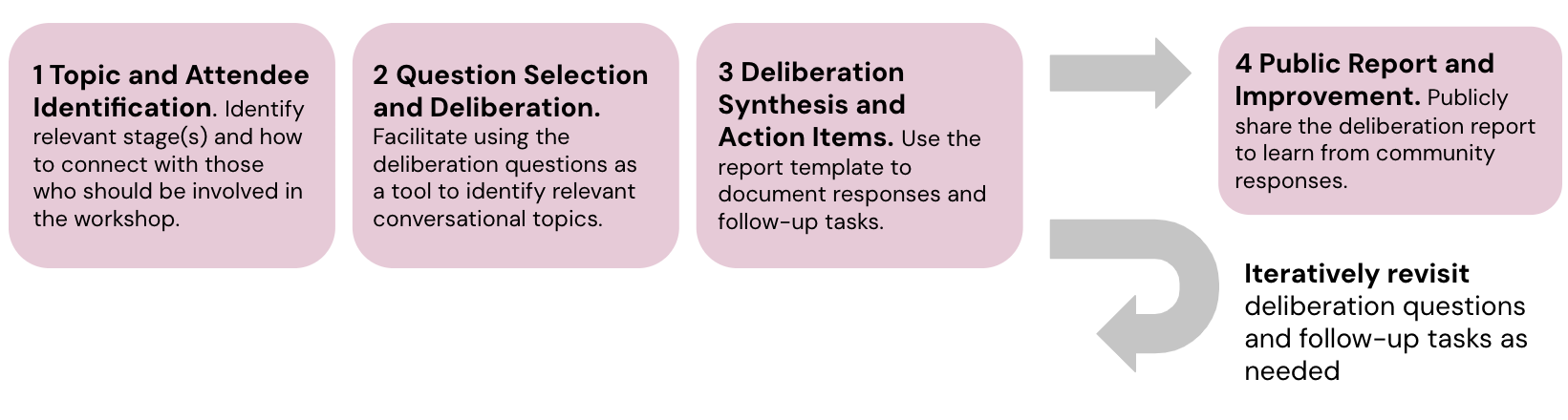}
    \caption{A high-level overview of the main stages involved in the proposed use case for the\textit{ Situate AI} Guidebook, intended to support structured and iterative deliberations within a public sector agency.}
     \label{fig:guidebook_process}
\end{figure*}

Participants envisioned that the guidebook could be effectively used to support structured, iterative deliberation through formal workshops between members of their agency. In this section, we elaborate on one possible way this use case can be instantiated into an overall deliberation process, then discuss how this compares to participants’ existing practices within their agency. We provide an example of one possible implementation of a formal deliberation process, using the guidebook. 


\paragraph{\textbf{Example instantiation of proposed deliberation process}}
This process involves a four-stage phase, where the public sector agency would first appoint a facilitator(s) to help organize the overall decision-making process:

\textbf{Stage 1: Topic and Attendee Identification. }The facilitator will identify which of the four guidebook topics, if not all, they are interested in convening a deliberation workshop on. Based on broad guidance provided in the guidebook, the facilitator will then identify the stakeholders that should be included in the deliberation workshop based on the selected topics. For example, the Societal and Legal Considerations section is designed to be used by a more diverse range of stakeholders (e.g., AI practitioners, frontline workers, community members, legal experts) compared to the Data and Modeling Constraints section (e.g., AI practitioners only). 

\textbf{Stage 2: Question Selection and Deliberation.} After finding a shared time for the deliberation workshop, the facilitator should share the goals and topics of deliberation (included in the guidebook) with the group. The guidebook includes a large number of questions for each topic of deliberation. For example, the \textit{Goals and Intended Use} section alone has 52 questions. To ensure that the questions can be feasibly discussed within the allocated time, we highlight 1-2 recommended questions per major subsection, resulting in a smaller number of questions (19 questions for the \textit{Goals and Intended Use }section). The remaining questions are also available in the guidebook as ``optional questions.'' We recommend that, at the start of the workshop, the facilitator provides the attendees with the opportunity to identify any questions from the ``optional questions'' category they would like to additionally or alternatively discuss in the workshop. As the attendees are discussing each question, the facilitator should take note of their responses and points of disagreement. If there are disagreements that are challenging to resolve in response to a question, the facilitator should help the group identify action items to help gather more information or perspectives and plan to revisit the question at a later time. If the group finds they currently lack the resources or knowledge to fully address a question, the facilitator should also make note of this and plan to revisit the question at a later time. 

\textbf{Stage 3: Deliberation Synthesis and Action Items.} After the deliberation workshop, the facilitator should summarize the \textcolor{black}{discussions and outcomes}\st{deliberations} into the deliberation report template which we include in the guidebook. The template includes the following questions: \textit{(1) What is your recommendation? (2) Please list core reasons for your recommendation, based on the deliberation workshop, (3) Are there any follow-up tasks you must complete, in order to fully support this recommendation? If yes, please write the task(s) and plan(s) for completing it, and (4) What core counter-arguments against this recommendation arose during the deliberation workshop? Please describe each counter-argument, including how you addressed or plan to address each one. }Based on the report responses, the facilitator should continue to iteratively revisit the deliberation questions, organizing additional deliberation workshops with the attendees as needed, as they complete the follow-up tasks included in the report. 

\textbf{Stage 4: Public Report and Improvement.} The facilitator should work with their agency to publicly share the deliberation report with agency-external stakeholders, including impacted community members and related organizations. To promote conversations and bidirectional learning \textcolor{black}{between}\st{from } community members \textcolor{black}{and agency-internal stakeholders}, the agency should hold public convenings and host an online commenting forum for any individuals who would feel more comfortable contributing anonymously online. The agency should then synthesize the themes that emerged from the conversations, identify action items to address any concerns, and share these results of the community conversations with the public. In the guidebook, we plan to include links to existing community review efforts to provide examples of what this interaction could look like. However, we note that effectively completing this step requires additional research and resource creation (which we discuss in the Discussion Section~\ref{discussion}).

\paragraph{\textbf{How participants' existing organizational practices informed the proposed process design}}
Reflected in the process above, participants raised several important considerations to ensure the process is practical and meaningful to their agency. For example, participants in agencies that are actively developing new AI tools described that there are already AI design and development processes in place that support focused discussions on improving, for example, the algorithmic fairness of their AI tool. These participants did not want the Situate AI Guidebook to replace these conversations and work sessions. Instead, they desired a process that could\textbf{ augment their existing processes}. For instance, participants noted that having these deliberation workshops earlier on, before they developed or analyzed any AI model, can help promote reflexive conversations about what it means to do the work of AI fairness and what important considerations that should aid this work (e.g., whether there is a definition of ``fairness'' that agency workers agree on). Relatedly, as reflected in the process description, several participants described the importance of \textbf{revisiting these deliberation questions iteratively}, throughout the AI development and deployment process (rather than only discussing these at the early ideation stages of a development process). For instance, participants described that their understanding of some of the question responses (e.g., the intended outcomes that the AI tool should help achieve) may evolve with time as the AI tool development evolves (e.g., depending on what is technically possible, given data constraints or prediction errors). Participants noted that designing the guidebook to center an iterative process would help ensure the conversations complement their existing AI development process, which is also iterative in nature. We originally designed the process so that the deliberation workshop attendees would be required to come to a consensus on the final recommendation before moving forward. However, the participants we spoke with emphasized that this may be impractical and unnecessary based on their end goals for the guidebook. We elaborate on this point in Section~\ref{indicators} when we discuss the guidebook’s Success \textcolor{black}{Criteria}. 

We discuss limitations and future work related to this proposed use case in the Discussion, drawing on prior literature suggesting ways in which research-based conversational tools may not be adopted in practice or may be used in inappropriate ways. 

\subsubsection{\textbf{Empowering Participation: Accounting for Organizational Power Dynamics.}} \label{process_design_power}
Findings from this study strongly suggest that frontline workers---those who would be asked to use the AI tool once deployed---are interested in engaging in early-stage deliberations around what the AI tool should be designed to assist. Prior literature also suggests that ensuring agency leadership and AI developers understand frontline workers' needs and challenges is critical to ensuring that the ``right'' AI tools are being developed \textcolor{black}{(e.g.,~\cite{kawakami2022improving,yang2019unremarkable,saxena2020human,kawakamistudying})}. However, effectively supporting conversations between roles with prominent and knowledge differentials remains a challenging task~\cite{HoltenMoller2020}. For the current version of the \textit{Situate AI} Guidebook, we begin accounting for this challenge by editing the language used in some of the guidebook sections to ensure it is understandable to those without prior knowledge on technology. We additionally asked frontline workers about their desires for how they would want to participate in the deliberation process, to ensure they feel safe to share any concerns. In this section, we discuss these findings. However, we note that future work is needed to ensure that the\textit{ Situate AI} Guidebook ($Ver.1.0$) adequately accounts for organizational power dynamics. In the Discussion, we discuss implications for complementary policy interventions that may be needed for an agency to effectively facilitate a multi-stakeholder deliberation process.  

\paragraph{\textbf{Frontline workers' preferred processes for participating in multi-stakeholder deliberations}}
Frontline workers in our study had a range of perspectives around how best to involve them and their colleagues in the deliberation process. One frontline worker suggested that their agency should require all frontline workers to attend the deliberation workshops. They described that, without making participation in these discussions mandatory, frontline workers may opt to skip meetings given their busy schedules. The participant further expressed concerns that, if participation was on a voluntary basis, frontline workers who join may be harmfully self-selective: ``... particularly for people with marginalized identities, it’s important for them to be a part of these spaces and voice their concerns. I think that if it wasn’t mandatory, it might be, you know, [a] self-selecting group'' (F7). This frontline worker, along with another worker, also described how adjustments to the proposed process could help frontline workers feel more comfortable raising concerns. For example, the participants expressed that, for multi-stakeholder conversations, some frontline workers may feel more comfortable having a separate frontline worker-only deliberation workshop, synthesizing and formalizing their perspectives, then going to a group meeting with other agency stakeholders to present their perspectives. As the participant described:
\begin{quote}
    ``A lot of social workers are very non-confrontational [...] we are our clients’ best advocates but not for ourselves. And so I definitely do think that people might be more comfortable, you know, having their own sort of peer group discussion or colleague group discussion. And then that being you know, sort of the concerns being written down and formalized, and that being taken rather than a more informal sort of like anyone who has concerns just raise their hand and say their piece. I feel like that might be a bit daunting for some people'' (F7). 
\end{quote}

\paragraph{\textbf{Alternative use case: Using the guidebook to empower everyday conversations}} \label{process_everyday}
Complementing frontline workers' desires for engagement, participants from agencies that were not yet developing new AI tools described a different use case for the Situate AI guidebook. These participants envisioned that the guidebook could be used by teams to support everyday conversations, with the aim of proactively avoiding pitfalls in AI project ideation and selection. For instance, one participant described that they wanted the guidebook to be used more casually, by everyone in the agency, to help all staff members feel empowered to ``be able to do a little more of the innovation'' (L7). The participant described that, even if they get stuck or need help, ``it would be awesome for them to have a library of resources that they can look at'' to help them get started. The participant further described that workers in their agency should ``have the flexibility to structure the deliberation workbook to their needs,'' for example, deciding which questions to discuss, how much time to take in discussing the questions, and who to talk with. By having a guidebook that empowers any staff member to discuss topics around AI, the participant hoped that these deliberations could have rippling effects on their agency’s overall culture: 
\begin{quote}
    ``Maybe we're trying to get from [...] 'let's do this big project right with the right leaders and things in the room' to 'how do we create a culture of improvement' [...] Not just how do we do a technology project the right way, but actually can it have a broader impact on culture. This is how we do anything. It's always with this batch of questions in mind and thinking about how we can be people around problem solving'' (L7).
\end{quote}

\paragraph{\textbf{Desires to expand participation to community members for future versions of the guidebook}} \label{process_comm}
Several participants across agencies and community advocacy groups were interested in involving community members in the deliberation workshops supported by the guidebook. Workers within the agency wanted guidance on how to do this effectively. In our conversations with community advocates, we probed on how they would like to be involved in deliberations around the Situate AI guidebook. They described the importance of compensating community members for their time, providing multiple channels for communication (e.g., online forums and in-person meetings), and following up with the outcomes of the conversations: ``that happens a lot, you know–agencies are like 'oh, we engage with the community, and we brought them into the space with us.' But then there's no follow up or follow through from those conversations. And that's been a historical thing” (C2). 

Importantly, we note that the guidebook, in its current form, is designed to support conversations across workers within a public sector agency. It is not designed to directly support conversations between agency-internal workers and agency-external stakeholders (e.g., community members). In the Discussion section, we discuss opportunities to expand participation through design improvements. 

\subsection{Success \textcolor{black}{Criteria}}\label{indicators}
What outcomes do public sector agencies and impacted community members consider ``meaningful,'' when assessing the effectiveness of the \textit{Situate AI} Guidebook? What are their underlying theories of change around how their public sector agency could move towards more responsible early-stage AI design practices, and how do they envision the \textit{Situate AI} Guidebook can help them progress towards that path? Overall, through the guidebook, participants wanted to form an understanding of the disagreements and tensions across agency workers felt most strongly about, to help position themselves to better address these disagreements through changes to the problem formulation or design of \textcolor{black}{a proposed AI tool}\st{the AI tool idea}. Importantly, as indicated by the process design in Section~\ref{process_design}, participants described that the goal of the Situate AI Guidebook should \textit{not} be to \textit{resolve} these tensions and disagreements across individuals. Participants described that this is an infeasible task, given underlying differences in values and goals across agency stakeholders. 
In this section, we elaborate on four success \textcolor{black}{criteria} of the guidebook. These success \textcolor{black}{criteria} intend to help communicate the intended goals and boundaries of the guidebook, and including how they are informed by participants’ own notions of success for deliberations around the design of AI tools. 

\subsubsection{\textbf{Make it easier for different agency stakeholders to communicate with each other about AI design, evaluation, and governance considerations.} }
Many of the challenges that participants in our study described could be addressed if better communication channels existed between different agency stakeholders---including amongst AI developers, agency leadership, and frontline workers. This challenge is also well-documented in prior literature studying public sector AI decision-making~\cite{saxena2020human,kawakami2022care,HoltenMoller2020,kawakamistudying}. For instance, in current practice, frontline workers are often not meaningfully involved in early-stage deliberations around AI design and adoption. As a result, agency leadership and AI developers have interpreted workers’ concerns around AI as a signal for not understanding what the AI tool does. Involving frontline workers in these earlier discussions can both help more proactively inform workers of AI capabilities and empower them to engage in constructive conversations that would improve the design of AI tools. 

\subsubsection{\textbf{Bring context-specific needs for resources to the forefront of AI project selection conversations.}}
While participants often knew which resources they needed to successfully implement a given AI tool, their past challenges sometimes surfaced a missed opportunity to identify these needs at an earlier stage of their AI design process. Moreover, related to \textcolor{black}{the previous success criterion}\st{Indicator-1 above}, our conversations with the participants surfaced ways in which having agency workers with different roles and perspectives engaged in these early-stage deliberations can strengthen their ability to anticipate the potential impacts of AI design decisions. For example, frontline workers voiced that AI tools they had used in the past were designed in ways that conflicted with their existing decision-making policies; other workers described that AI deployments may add additional labor to their day-to-day tasks, given they may be asked to more diligently collect data. In current AI development processes, where frontline workers may only be meaningfully engaged in AI implementation or piloting stages, mitigating these negative impacts may require more substantive tasks like redesigning the AI tool. Participants further described that these resource-related needs were highly context-specific. For instance, when discussing the importance of anticipating how their AI tool may impact community members, one participant recalled how even the definition of “community” may differ across agencies and AI tools: “Because we would always say, 'we're doing stuff [where] the community is informing us.' And then we realized, 'oh wait, it wasn't necessarily the group of people who were impacted by [our decisions]'” (L7).

\subsubsection{\textbf{Make social and ethical considerations a first order priority in conversations around whether to move forward with an AI tool idea.}}
As described in Section~\ref{topic_social} and~\ref{topic_org}, participants described their past assessments around whether an AI tool was appropriate to implement centered algorithmic considerations---whether that be the quality of their training data or outcomes of algorithmic fairness or accuracy metrics. While these considerations are critically  important, others also discussed a desire to rigorously deliberate the underlying values embedded in design decisions, and the social and ethical impacts of a proposed AI tool. This echoes concerns from prior literature, discussing how technical design decisions often include hidden policy decisions and value judgements~\cite{gerchick2023devil,stapleton2022imagining}. Prior work also suggests the importance of these considerations, noting that existing AI ethics toolkits have largely framed the work of ethics as ``technical work''~\cite{wong2023seeing}.

\subsubsection{\textbf{Make “fitting” an AI tool into a workplace a design problem, rather than an implementation problem.}}
This success \textcolor{black}{criteri\textcolor{black}{on}\st{a}} intends to avoid practices where the AI tool idea is conceived before fully understanding context-specific practices and needs, and in turn, creating AI tools that frontline workers must then attempt to ``fit'' into their existing workflow. This tendency to treat ``fitting'' AI tools as an implementation problem, and its negative impacts on workers’ ability to improve their existing decision-making practices, is also well documented in prior literature\textcolor{black}{~\cite{yang2019unremarkable,yildirim2023creating}}. Indeed, participants---including both AI developers and frontline workers-described that they wished they could have had better conversations, early on, to understand what the actual goal of the tool they were building should be. As one AI developer described, recalling a past experience in their team where leadership had asked them to create an AI tool: “It was kind of hard to get a sense of what the actual issue was that was being asked to be solved. It sounds kind of a lot like, `Here's a bunch of different potential places an algorithm might fit in’” (A04). Ultimately, they were asked to create a predictive model to “to find fraud where there wasn't already suspicion of fraud.” However, the AI developer described feeling leadership had proposed the idea as ``this cool thing we could do'' but, in reality, realizing that creating such a tool would create more problems downstream in the system (in this case, it would create too many referrals to be able to investigate). By promoting early-stage, structured deliberations around critical topics related to AI tools, public sector agencies could be supported in identifying \st{more under-explore\textcolor{black}{d,}\st{ and}}higher-value\textcolor{black}{, lower-risk} opportunities to innovate with AI systems.

\section{Discussion}\label{discussion}
Public sector agencies in the U.S. are increasingly exploring how new AI tools can assist or automate services in child welfare, homelessness housing, healthcare, and policing, among other \textcolor{black}{domains}~\cite{chouldechova2018case,green2020false,yang2019unremarkable,obermeyer2016predicting,saxena2020human}. In the U.S., these public \st{sector }services have historically been characterized by racial inequity, procedural injustice, and distrust from the \textcolor{black}{impacted} communit\textcolor{black}{ies}~\cite{eubanks2018automating,brown2019toward,stapleton2022imagining}. While agencies have rapidly begun to deploy AI tools to improve \textcolor{black}{their} services, ensuring responsible development has proven to be an immense challenge. In the past decade, \textcolor{black}{such AI tools} have often failed to serve the needs of the communities that agencies are expected to serve~\cite{eubanks2018automating,bao2022artificial,gerchick2023devil,ho2022algorithm,ho2023algorithm}. A growing body of literature has recognized that many downstream harms resulting from AI tools can be traced back to decisions made during the earliest problem formulation and ideation stages of \st{AI design}\textcolor{black}{the AI lifecycle}. Yet, there are few, if any, effective resources for public sector agencies in making more deliberate decisions regarding \textit{whether} a given AI \st{tool}\textcolor{black}{proposal} should be developed in the first place.

Through iterative co-design sessions with 32 individuals (agency leaders, AI developers, frontline decision-makers, and community advocates) across four public sector agencies and three community advocacy groups, we created the \textit{Situate AI} Guidebook ($Ver.1.0$). The guidebook, designed \st{to be used by}\textcolor{black}{for} public sector agency workers, scaffolds \textcolor{black}{the} process for early-stage deliberations around \textit{whether and under what conditions} to move forward with the implementation or adoption of a new AI tool \textcolor{black}{or} idea. To support this process, the guidebook presents a set of 132 deliberation questions--which participants indicated are critical to consider yet \textcolor{black}{are often} overlooked today--spanning \st{across} both social (organizational, societal, and legal) and technical (data and modeling) considerations around AI; along with guidance on the overall deliberative decision-making steps and success criteria for use. In this section, we discuss the design decisions we made in creating the guidebook, along with limitations and opportunities for future work. For each section \textcolor{black}{of this discussion}, we begin by summarizing relevant portions of the findings. Then, we elaborate on limitations and future opportunities to improve upon the existing guidebook. 

\subsection{Overcoming Low Adoption Rates for Responsible AI Toolkits in the Public Sector}
As the research community continues to innovate new Responsible AI toolkits, recent literature has raised concerns regarding the \textcolor{black}{practical} efficacy of such toolkits. Prior work \st{examining responsible AI toolkits have}\textcolor{black}{has} found that the majority of AI ethics toolkits fail to account for the \textcolor{black}{relevant} organizational context\st{ in which it is designed for, raising concerns around}\textcolor{black}{, hindering} their usability (e.g., overlooking guidance on \textit{how} different stakeholders should be engaged) and effectiveness (e.g., focusing on the technical but neglecting the social aspects of AI ethics work) \textcolor{black}{~\cite{wong2023seeing}}. \textcolor{black}{Public sector decision-making around service allocation is often shaped by resource and staffing shortages, and require balancing tradeoffs to meet the competing needs of a range of stakeholders (e.g., impacted community members, policymakers and regulators, politicians)~\cite{veale2018fairness}. Moreover, AI tools in the public sector often target socially high-stakes decisions (e.g., whether to screen in a family for child maltreatment investigation, or provide an individual with a credit loan), that have disproportionately negatively impacted the lives of historically marginalized communities.} \textcolor{black}{Prior work has shed light \textcolor{black}{on} the downstream impacts that public sector AI systems have had (e.g.,~\cite{brown2019toward,robertson2021modeling}), along with challenges to ensuring their responsible design and use (e.g.,~\cite{saxena2020human,kawakami2022improving,veale2018fairness}). Through our study, we demonstrated how collaborating with public sector agencies and community members to co-design \textcolor{black}{a} responsible AI toolkit--including its process and content design--can help surface and account for some of these challenges. That said, future research is needed to understand how effective the toolkit is in practice, and to surface other challenges that can only be observed through \st{in-practice}\textcolor{black}{actual} use (rather than through our co-design and interview study format). In the following subsections, we briefly discuss related findings and opportunities \st{for future work} to improve the contextual design and use of the Situate AI Guidebook for public sector settings.} 

\subsubsection{Designing for more inclusive forms of worker participation.} 
\textcolor{black}{While AI tools for public sector contexts implicate a range of different agency-internal stakeholders, these agency workers---from agency leaders to frontline workers---often operate in silos, separated by power imbalances and knowledge differentials.} \textcolor{black}{We found that participants desired a range of particiation structures to account for these differences.}\st{To understand how we can account for these relational dynamics when different agency workers participate in the deliberation process, we asked frontline workers about what interventions (e.g., organizational policies, alternative forms of participation or policies) they would find useful. Participants’ responses pointed to unexpected use cases for the toolkit.} For example, some frontline workers wanted to first gather amongst others with similar occupations to prepare for the deliberation workshop, and then send in one frontline worker to attend the workshop and represent their perspectives. On the other hand, other participants suggested that there should be an organizational policy that required all frontline workers to attend the deliberation workshops alongside the AI developers and agency leaders. Future work is needed to understand the broader range of solutions that could best address these differences in workers’ preferences.  
For example, future work could pilot different processes, where agency workers are grouped in deliberation workshops in specified configurations depending on their role and background.\st{Through observations and retrospective interviews of these piloted processes, we can learn whether certain configurations of role-based groupings help workers in lower positions of power feel safer when raising their concerns. Moreover, piloting different processes can also help us understand the potential tradeoffs of having workers of similar or different backgrounds grouped together in a workshop. } Through observations and retrospective interviews of these \textcolor{black}{configurations}\st{piloted processes}, we could better understand whether\st{, in the context of discussing normative or ethical concerns regarding a given AI design concept,} having a set of deliberation questions alone is adequate to prompt meaningful conversations\st{ across workers with similar or different backgrounds}. Future work could additionally explore how additional resources and tools
could be used alongside the deliberation toolkit, in order to effectively scaffold conversations around the appropriateness of AI design ideas. This direction would be especially critical to pursue, in order to ensure that the deliberation toolkit is accessible to those who may not have had prior exposure to AI technologies.

\subsubsection{Incentivizing and governing responsible use}
\st{The range of use cases that participants envisioned also sheds light onto an important tradeoff between the flexibility and efficacy of a designed responsible AI toolkit. For example, while a toolkit that intends to ``empower everyday conversations’’ (Section~\ref{process_everyday}) may be desirable for some agencies, it may also be the case that having a toolkit ``with teeth’’ (e.g., requiring agencies to use the toolkit to support formal, structured workshops) may be more likely to lead to substantial improvements.}
Ensuring responsible use and adoption of the \textit{Situate AI} Guidebook may require complementary efforts from governing bodies. For example, while the U.S. does not currently require public sector agencies to \st{communicate with frontline workers when designing AI tools}\textcolor{black}{document early-stage deliberations around AI}, having similar forms of external forces that incentivize agencies to engage in early-stage deliberation may help ensure that the deliberation toolkit is used effectively. 
\st{Besides creating new government regulation, o}One way to incentive public sector agencies may be to clearly communicate how the toolkit aligns with and complements existing voluntary guidelines, such as those in the NIST AI Risk Management Framework (RMF)~\cite{nistRMF}. While the NIST RMF and NIST RMF Playbook~\cite{nistRMFPlay} both focus on providing higher level guidance on steps to follow for responsible AI design, research-based co-designed toolkits like the Situate AI Guidebook can help bridge gaps between their proposed policy \textcolor{black}{guidance} and real-world practice. In future work, we plan to map the guidebook to the four functions captured in the AI RMF Core: Govern, Map, Measure, and Manage. For example, each question or category of questions could be assigned one or more of the AI RMF Core functions. 

In future work, we plan to explore with public sector agencies \textcolor{black}{community advocates, and other stakeholders}\st{ themselves} how new policy and organizational interventions can support them in using the \textit{Situate AI} Guidebook. The public sector agencies in our study, including the frontline workers, expressed interest in exploring how to use the guidebook in practice \textcolor{black}{through pilots}. 

\subsection{Expanding the \textit{Situate AI} Guidebook to Engage Community Members}
\textcolor{black}{In public sector contexts, there is often a greater expectation that decisions center the needs of the community, including by being transparent to and engaging with the community during the decision-making process}. In our study, participants expressed a desire for guidance on how to engage with community members in discussing complex topics around AI design. While the deliberation protocol is not currently designed to support such conversations, the current version poses questions that suggest follow-up tasks involving conversations with community members. For example, the question ``Are we assessing the tool from the perspective of impacted community members? What evidence do we have to suggest that we are genuinely understanding their concerns and desires?'' suggests that the agency should talk with impacted community members to understand their perspectives--a task that would require additional guidance and resources to complete successfully. Agency workers acknowledge they are often pushed to involve community members in their AI design work but without actionable guidance on how to do so effectively. Participants suggested linking existing relevant resources from the guidebook, to assist agencies in \textcolor{black}{this regard}. Moreover, community advocates in our study expressed interest in engaging in the deliberation workshops themselves. 

Future work should explore ways to improve the design, structure, and process of the deliberation guidebook so that it is well-equipped to support conversations between agency-internal workers and agency-external community representatives. For example, to help bridge a shared vocabulary about AI between agency workers and community representatives, future work could begin by integrating existing resources and guidance from publicly available guides like \emph{A People’s Guide to Tech}~\cite{peopleAI}. It is possible that the specific questions and topics this deliberation guidebook addresses requires additional scaffolding and support. Future work should explore ways to provide this support through continued collaborations with community advocacy groups. Community advocates in our study additionally expressed interest in having both the option to attend in-person workshops and to participate anonymously online. Future work could explore ways to support more democratic forms of participation online using social computing platforms \textcolor{black}{(e.g.,~\cite{kriplean2012supporting})} intended to facilitate and analyze deliberation about specific topics around AI\st{(e.g.,~\cite{})}.

\subsection{Exploring how the \textit{Situate AI} Guidebook Can Support Deliberation in Non-Public Sector Contexts}
While the \textit{Situate AI} Guidebook was originally designed for high-stakes public sector decision-making domains, there is an opportunity to adapt it to meet the needs of other \textcolor{black}{AI use cases}\st{real-world AI deployment contexts}. Private and public sector settings share many organizational \st{and worker} challenges (e.g., communication barriers across teams and occupations) and development tendencies (e.g., targeting problem spaces that AI capabilities may be ill-suited towards), that could implicate the effective design of responsible AI toolkits. Moreover, by designing for a setting with relatively high expectations and standards for responsible design (i.e., \textit{high stakes} AI applications in the \textit{public sector}), the \textit{Situate AI }Guidebook sets a high bar for the kinds of questions and processes that should be followed to responsibly evaluate early-stage AI design concepts \textcolor{black}{elsewhere}. For this reasons, we expect that the guidebook may \st{also have some unique strengths that make it}\textcolor{black}{be (at least partially)} applicable to other AI deployment contexts, including certain high risk applications in industry (e.g., healthcare, credit lending). 

Indeed, many of the questions that the participants generated are relevant to AI deployment contexts beyond the public sector. \textcolor{black}{Most} deliberation questions\st{, although including some questions that may be uniquely tailored to the public sector, overall} target core issues relevant to all AI deployments (i.e., around the goals, ethical implications, technical constraints, and governance practices surrounding an AI deployment). Besides the deliberation questions, the design principles underlying the guidebook may also help make the guidebook useful for non-public sector contexts. Because the public sector agencies we partnered with differed in their organizational practices (e.g., who is involved in decision-making around AI) and priorities (e.g., types of services provided), we intentionally designed the guidebook to allow for flexible adoption and personalization. 
For instance, in designing towards Design Principle 2, we categorized the questions into modular topics and subtopics that can be selected and combined for a given deliberation workshop. There is an opportunity for future work to expand the set of labels attached to the deliberation questions. For example, participants described an interest in future versions of the toolkit that categorized questions based on the type of technology (e.g., generative AI, predictive analytics)\st{that the question is most relevant for}, or the type of deliberation (e.g., individual assumption-checking, knowledge-sharing, future task identification) that the question is intended to support. Future work could similarly aim to understand the types of questions that are the most critical for certain AI deployment contexts (e.g., public sector social work, private sector healthcare, etc.).

\section{Conclusion}
As public sector agencies in the U.S. increasingly turn to AI tools to increase the efficiency of their services, it becomes critical to ensure these tools are designed responsibly. While much research and development efforts have been dedicated to better scaffolding responsible AI development and evaluation practices, real-world AI failures often point to fundamental problems in the problem formulation of an AI tool--problems that should be addressed before proceeding with any decision to develop an AI tool. Yet, we currently lack effective processes to support such early-stage, deliberate decision-making in the public sector. This paper introduces the \textit{Situate AI} Guidebook: the first toolkit that is \textit{co-designed} with public sector agencies and community advocacy groups to scaffold \textit{early-stage deliberations} regarding \textit{whether or not} to move forward with the development of an AI design concept. Through co-design sessions conducted over the course of 8 months, participants generated 132 questions which we organized under four high-level categories including (1) \textit{goals and intended use}, (2) \textit{social and legal considerations} of a proposed AI tool, (3) \textit{data and modeling constraints}, and (4) \textit{organizational governance factors}. In this paper, we elaborate on how participants' practices, challenges, and concerns shaped the \textit{Situate AI} Guidebook's guiding design principles, the deliberation questions they believed were critical for early-stage decision-making, the overall organizational and team decision-making process the guidebook should scaffold, and the success criteria used to assess the effectiveness of the guidebook. We additionally discuss opportunities for future work to improve the design and implementation of the \textit{Situate AI} Guidebook, including via continued partnership with public sector agencies in our study, who plan to pilot how the guidebook can be used in their agency.

\begin{acks}
We thank the public sector agencies and community advocacy groups that shared their time, perspectives, and expertise with us to help create this paper and guidebook. We also thank the anonymous reviewers for their thoughtful feedback that helped improve this paper. The researchers gratefully acknowledge the support of the Digital Transformation and Innovation Center at Carnegie Mellon University sponsored by PwC. This research was also supported by funding from the UL Research Institutes (through the Center for Advancing Safety of Machine Intelligence), the National Science Foundation (NSF) (Award No. 1952085, IIS2040929, and IIS2229881), the NSF Graduate Research Fellowship Program, and the K\&L Gates Presidential Fellowship (through Carnegie Mellon University). Any opinions, findings, conclusions, or recommendations expressed in this material are those of the authors and do not reflect the views of NSF or other funding agencies.
\end{acks}

\appendix
\section{Deliberation Questions} \label{appendix}
This section includes the full list of questions included in the \textit{Situate AI} Guidebook\footnote{\url{https://annakawakami.github.io/situateAI-guidebook/}} ($Ver.1.0$).
\subsection{Goals and Intended Use}
The set of questions below are intended to support conversations around the following broader question: \textbf{Given our underlying goals and intended use case(s), is our proposed AI tool appropriate?} This stage would benefit from the expertise of the following stakeholders at the minimum, amongst others: Agency leadership, AI practitioners, frontline workers, community members.  
\subsubsection{\textbf{Overall goal for using algorithmic tool}}
\paragraph{Who the tool impacts and serves}
\begin{itemize}
    \item Who is going to be affected by the decision to use this hypothetical AI tool?
    \begin{itemize}
        \item Who is going to be the most impacted?
    \end{itemize}
    \item Who benefits from the use of the tool?
    \begin{itemize}
        \item To what extent are the targeted outcomes intended to benefit the agency, versus the community?
    \end{itemize}

\end{itemize}

\paragraph{Intended use}
\begin{itemize}
    \item What evidence do we have suggesting that the painpoint this tool aims to solve actually exists?
    \item What evidence do we have suggesting that technology may offer a remedy to this painpoint? (Evidence may include, for example, historical agency metrics, legislature, community members, research reports.)
    \begin{itemize}
        \item What evidence suggests the specific form of technology we are envisioning (e.g., predictive analytics) may offer a remedy?
    \end{itemize}
    \item What are the additional challenges and risks associated with pursuing a technological solution to this problem? 
\end{itemize}

\paragraph{Involving agency-external stakeholders in determining the goals}
\begin{itemize}
    \item Think about the most impacted stakeholders you identified in response to the questions above. How do we bring their voices to the table when determining goals? 
    \item How can we open opportunities for those who are most impacted by the new tool to inform the decision-making process? 
    \item When will we start to engage impacted communities in discussions around how the tool should be designed or used?  
\end{itemize}

\paragraph{Differences in goals}
\begin{itemize}
    \item Are there differences in the goals the agency versus community members think the tool should address? If so, what are they? If we are uncertain, what can we do to understand potential differences? 
    \item What evidence do we have that we adequately understand the outcomes the community cares about? 
    \item To what extent are we optimizing the things the agency cares about versus what impacted community members care about?
    \item Is the process we have in mind for achieving a community-oriented outcome (e.g., child safety) also aligned with the community's desires?
\end{itemize}

\paragraph{Envisioned harms and benefits}
\begin{itemize}
    \item What are the potential harms and benefits of the tool, and to whom?
\begin{itemize}
        \item Do benefits outweigh the harms? 
        \item Do we expect there to be tradeoffs between accuracy, fairness, explainability? For example: making decisions in a completely random fashion may look “fair”, but is not necessarily accurate.
        \item Will this tool help us better allocate (scarce) resources?
\end{itemize}
    \item What biases (as a government agency) do we bring into this decision-making process?
    \begin{itemize}
        \item How can we identify and mitigate them? What forms of collaboration (e.g., with impacted community members) can help us do this?
    \end{itemize}
    \item How does this tool help us better deliver to the people we are serving, if at all?
\end{itemize}

\subsubsection{\textbf{Selection of outcomes that the algorithmic tool aims to improve}}
\paragraph{Impacts of outcome choice}
\begin{itemize}
    \item Hypothetically, imagine that our tool does a perfect job of improving the outcome that it targets. What additional problems might this create elsewhere in the system?
    \item To what extent are we optimizing the things the agency cares about, versus what impacted community members care about?
\end{itemize}

\paragraph{Assumptions behind outcome choice}
\begin{itemize}
    \item What assumptions are we making, when deciding what the tool should optimize?
    \item How are we operationalizing goals for the tool, e.g., improving child 'safety'? What assumptions are we making?
    \item How do we bring providers to the table to decide on the use of outcomes?
\end{itemize}

\paragraph{Predictability of outcomes}
\begin{itemize}
    \item Have we run any tests on historical data records, to check whether we get predictions on this outcome that actually make sense?
    \item How rare is the event we are trying to predict? If it is rare, how reliably do we think we can predict it?
    \item How does the inclusion of additional information (e.g., attributes) improve outcomes?
\end{itemize}

\subsubsection{\textbf{Empirical evaluations of algorithmic tool}}
\paragraph{Measuring improvement based on outcomes}
\begin{itemize}
    \item Once the tool is deployed and in use, how can we evaluate how well it is working in the short-term? How can we evaluate how well it is working longer-term? 
    \item What are some ways we might evaluate whether this tool is successful in improving the targeted outcomes?  
    \item For evaluating worker-ADS decisions post-deployment: Do the decisions change by worker experience, worker demographic, or by supervisor?
    \item What performance measures do we plan to use to evaluate the tool? 
    \item What performance measures have already been used in early analyses of historical data, prior to the deployment of the tool? 
    \item Does this tool improve outcomes? How are we operationalizing "improve"?
    \item How does the use of the tool compare with the status quo? E.g., can we demonstrate the tool improves outcomes for the population of interest?
\begin{itemize}
        \item What is the "performance" and "fairness" of the existing baseline/status-quo decision process?
        \item Is there someone with relevant domain expertise that could help explain anomalies or trends?
\end{itemize}
    \item Do we think there are tradeoffs between accuracy, fairness, explainability? If so, what are they?
    \item How are we measuring negative and positive impact on families?
    \item Is there someone with relevant domain expertise that could help explain anomalies or trends?
\begin{itemize}
        \item How well do you understand the domain application of the historical data used in evaluation?
        \item Are there changes in policies and domain-specific practices in the historical data?
\end{itemize}
    \item Are there measured improvements resulting from the model's deployment?
    \item Are we using appropriate evaluation methods, e.g., synthetic controls, discontinuity analysis when cutoffs on risk exist.
    \item What outcome measures are we evaluating on? What can these measures tell us, and what can they not tell us?
\end{itemize}

\paragraph{Centering community needs}
\begin{itemize}
    \item How can we effectively evaluate the tool from the perspective of impacted community members?
    \begin{itemize}
        \item E.g., what does false positive, false negative mean for different impacted communities? How are we weighting false positives and false negatives, in a given use case, based on the relative costs of each type of error for the impacted stakeholders?
    \end{itemize}

\end{itemize}

\paragraph{Worker perceptions}
\begin{itemize}
    \item How might front-line workers respond to the tool? How can we better understand their underlying concerns and desires towards the tool? 
    \item How do front-line workers perceive the algorithm? (e.g., do they consider it a top-down requirement or a useful tool)
    \item Do domain experts also believe the model 'makes sense', e.g., selection of important features?
\end{itemize}

\subsection{Societal and Legal Considerations}
The set of questions below are intended to support conversations around the following broader question: \textbf{Given the societal, ethical, and legal considerations and envisioned impacts associated with the use of AI tools for our stated goals (identified in Facet 1), is our proposed AI tool appropriate?} This stage would benefit from the expertise of the following stakeholders at the minimum, amongst others: AI practitioners, frontline workers, community members, legal experts.

\subsubsection{\textbf{Legal considerations around the use of algorithmic tool}}
\begin{itemize}
    \item Do the people impacted by the tool have the power or ability to take legal recourse?
    \item Is there clarity around policies, e.g., whether algorithmic outcomes are included under 'public records'? 
    \begin{itemize}
        \item If someone asks for information around the tool, but there's no precedent, does the agency know what to do?
    \end{itemize}
    \item Are you having conversations with the Department of Justice and attorneys, to make sure the new decision models you implement will follow existing policies, procedures, statutes, and rules?
    \begin{itemize}
        \item Do you know which design decisions will be dictated by the law? For example: In the context of child maltreatment screening, if certain conditions are present in a case, then it is legally required to screen in for investigation.
    \end{itemize}
    \item Can you inform existing policies, procedures, statutes, and rules to better meet the needs of new decision models?
    \item Do you need a new temporary rule to receive permission to use the model?
    \item How are you interpreting challenges to ambiguities in prior legal decisions around the use of the tool?
    \item What are challenges to interpreting legal documentation and guidelines?
\begin{itemize}
        \item How well can we interpret case-specific considerations in the context of legal documentation/guidelines (e.g., when there is a lot of grey in practice, but the law is written in black and white)?
        \begin{itemize}
            \item E.g., in child maltreatment: "threat of harm" or "physical abuse" allegation type sounds black/white but there are various factors that make this grey. E.g., how hard did it hit them? Did it leave a mark? Action occurred but no impact from the action?
        \end{itemize}

\end{itemize}

\end{itemize}

\subsubsection{\textbf{Ethical and fairness considerations around the use of algorithmic tool}}
\paragraph{Impacted Community Member Needs}
\begin{itemize}
    \item Are there differences in the goals the agency versus community members think the tool should address? If so, what are they? If you are uncertain, what are your plans for understanding potential differences? 
    \begin{itemize}
        \item What are the envisioned harms and intended benefits from the tool that impact the community and the agency?
    \end{itemize}
    \item Can we have impacted community’s representatives or advocates at the table, to inform the design and use of the tool?
    \item How well are we engaging people closest to the problem and those impacted through the entire design, development, implementation, maintenance process?
    \item Are the outcomes intended for agency or community benefit?
    \item How well do we understand what outcomes the community wants to improve?
    \item Do we understand how impacted stakeholders perceive each decision? E.g., emotional valence, potential impacts, etc.
    \item To what extent are we optimizing the things the agency cares about versus what impacted community members care about?
\end{itemize}

\paragraph{Involving Impacted Communities}
\begin{itemize}
    \item What are underlying assumptions that tool developers/researchers may have, regarding the soundness of the design decisions made in the tool?
    \item How can we set up external participation opportunities, to increase access?
\begin{itemize}
        \item E.g., avoiding scheduling during a 9-5pm period (to open involvement to those who want to be involved)
        \item E.g., is it possible to involve groups that are not involved and paid by the agency, to get input and feedback?
        \item Do we know who should be included? How can we build the right network of people to talk with?
\end{itemize}
    \item Who has a seat at the table, to decide how the tool impacts you?
    \item How are you engaging with people closest to the problem (e.g., frontline workers,  community members, or others impacted by the decisions)?
    \item Have you communicated the limitations and historical context of the data, to community members?
    \item How well do we understand the costs, risks, and effort required of community members, if we invite them? E.g., many were directly harmed by decisions made by the agency.
    \item When do we start to engage impacted communities into discussions around the design or use of the tool?  
\end{itemize}

\paragraph{Clarity of Ethics Goals and Definitions}
\begin{itemize}
    \item Can we agree on a definition of fairness and equity in this context? What would it look like if the desired state is achieved?

\end{itemize}

\paragraph{Operationalization of Ethics Goals}
\begin{itemize}
    \item Are fairness and equity definitions and operationalizations adequately context-specific? (For example, in the child welfare domain: children with similar profiles receive similar predictions irrespective of race?)
    \item Do we know how to appropriately operationalize our fairness formulation in the algorithm design?
    \item Can we mitigate biases in the model?
    \item How can we balance tradeoffs between false negatives and false positives?
    \item How well are we integrating domain-specific considerations into the design of the tool?  
    \item Have we recognized and tried to adjust for implicit biases and discrimination inherent in these social systems that might get embedded into the algorithm?
\end{itemize}

\paragraph{Envisioning Potential Negative Impacts}
\begin{itemize}
    \item Do we understand the negative impacts of the decision made across sensitive demographic groups?
    \item What are the externalities / long-run consequences of the decisions?
\end{itemize}

\subsubsection{\textbf{Social and historical context surrounding the use of algorithmic tool}}
\begin{itemize}
    \item Have we recognized and tried to adjust for implicit biases and discrimination inherent in these social systems that might get embedded into the algorithm?
    \item How might we clearly communicate the limitations and historical context of the data to community members?
    \item Are you modeling historical, systemic patterns?
\end{itemize}

\subsection{Data and Modeling Constraints}
The set of questions below are intended to support conversations around the following broader question: \textbf{Given the availability and condition of existing data sources, and our intended modeling approach, is our proposed AI tool appropriate?} This stage would benefit from the expertise of the following stakeholders at the minimum, amongst others: AI practitioners.

\subsubsection{\textbf{Understanding data quality}}
\begin{itemize}
    \item How does the data quality and trends compare with an 'ideal' state of the world?
    \begin{itemize}
        \item What does our data look like, in terms of different demographic outcomes?
    \end{itemize}
    \item Has the definition of the data changed over time? (E.g., in child welfare, has reunification always meant to reunify with the parent?)
    \item What data do we have access to?
    \begin{itemize}
        \item Do we have the data/feature set to replicate the tool/analysis/and predictive accuracy of the existing tool?
    \end{itemize}
    \item How well do we understand the meaning and value of the data that will be used to train an algorithm?
    \item How is the quality of this data?
\begin{itemize}
        \item How accurate is the data?
        \item How recent is the data?
        \item How relevant is the data?
        \item Has the data been consistently collected?
\end{itemize}

\end{itemize}

\subsubsection{\textbf{Process of preparing data}}
\begin{itemize}
    \item How are we preprocessing the data?
    \item Who should be involved in making decisions around whether to include or exclude certain data points or features? Do we have plans for involving those people? 
    \item How do we address bias in the data?
    \item Do we have metrics for feature importance, that we could show relevant domain experts?
    \item How well do we understand the data collection process?
    \item Data leakage questions: Are we preventing oversampling of certain populations?
    \begin{itemize}
        \item E.g., in child welfare: Are we pulling one child per report, and one report per child, to ensure there's no information leakage between training and test sets?
    \end{itemize}

\end{itemize}

\subsubsection{\textbf{Model selection}}
\begin{itemize}
    \item Is our model appropriate given the available data? Why or why not?
\end{itemize}

\subsection{Organizational Governance Factors}
The set of questions below are intended to support conversations around the following broader question: \textbf{Given our plans for ensuring longer-term technical maintenance and policy-oriented governance, do we have adequate post-deployment support for our proposed AI tool?} This stage would benefit from the expertise of the following stakeholders at the minimum, amongst others: Agency leaders, AI practitioners, frontline workers.

\subsubsection{\textbf{Long-run maintenance of algorithmic tool}}
\paragraph{Measuring changes in model performance over time}
\begin{itemize}
    \item Do we expect there will be shifts in performance metrics over time? If so, why? What are our plans for identifying and mitigating those shifts? 
    \item Do we expect that the data collection process will improve over time? What might this imply for how we maintain the tool? E.g., Is there a need for adjusting thresholds over time?
\end{itemize}

\paragraph{Mechanisms to identify long-run changes}
\begin{itemize}
    \item Are we repeating feature engineering efforts over time?
    \begin{itemize}
        \item Are we detecting how trends shift over time at the population level?
    \end{itemize}
    \item Are there mechanisms in place that track whether certain data features have changed over the years?
    \item Do we have mechanisms to track longer-term outcomes over time, so that we can monitor for changes in model performance?
    \item Do we have the mechanisms to monitor whether the tool is having unintended consequences?
\end{itemize}

\subsubsection{\textbf{Organizational policies and resources around the use of algorithmic tool}}
\paragraph{Policies around worker interactions}
\begin{itemize}
    \item Is there training for frontline workers who will be asked to use the tool? What evidence suggests that this training is adequate? 
    \item How are frontline workers trained? 
    \item Is it clear to workers what information the tool can access, and what information it cannot?
    \begin{itemize}
        \item How is this communicated to workers?
    \end{itemize}

\end{itemize}

\paragraph{Governance structures}
\begin{itemize}
    \item Imagine that we could assemble the “ideal team” to monitor and govern the tool after it is deployed: What are the characteristics of this ideal team? 
\begin{itemize}
        \item Who is the \textit{actual} team that will monitor and govern the tool after it is deployed? 
        \item Given the gaps between the “ideal team” and the actual team we expect to have: What risks to post-deployment monitoring and governance can we anticipate? How might we mitigate these risks? 
\end{itemize}
    \item Are there appropriate forms of governance, around the implementation?
    \begin{itemize}
        \item Do those involved in governance have domain knowledge in the application context and have knowledge of the implementation process?
    \end{itemize}
    \item Are there sufficient guardrails in place to ensure algorithms wouldn't get weaponized?
    \begin{itemize}
        \item E.g., IRB-like programs and researchers at the same table, to minimize risk of weaponizing?
    \end{itemize}

\end{itemize}

\subsubsection{\textbf{Internal political considerations around the use of algorithmic tool}}
\begin{itemize}
    \item How well do we understand system administrators' and leadership's perspectives around the use of this tool?
    \item How well do staff and leadership understand 'why' the tool could bring value?
    \item Do system administrators and leadership perceive this tool positively?
    \item Do leadership support the future use of the tool?
    \begin{itemize}
        \item Do we have backing at a leadership level? E.g., director, agency, governor, community partners?
    \end{itemize}
    \item Is there sufficient buy-in from middle managers and executive support?
    \item Do we have mechanisms to address concerns that could come up during the ideation and design process?
\end{itemize}

\bibliographystyle{ACM-Reference-Format}
\bibliography{situateAI}


\end{document}